\input harvmac


\input epsf

\newif{\ifeq}           
\eqfalse               
                                %
                                %

\newcount\figno
\figno=0
\def\fig#1#2#3{
\par\begingroup\parindent=0pt\leftskip=1cm\rightskip=1cm\parindent=0pt
\baselineskip=11pt \global\advance\figno by 1 \midinsert
\epsfxsize=#3 \centerline{\epsfbox{#2}} \vskip 12pt {\bf Figure
\the\figno:} #1\par
\endinsert\endgroup\par
}
\def\figlabel#1{\xdef#1{\the\figno}}


\def\square{\kern0.5pt\vbox{\hrule height 0.8pt\hbox{\vrule width 0.5pt
\hskip 3pt \vbox{\vskip 6pt}\hskip 3pt\vrule width 0.3pt}\hrule
height 0.3pt}\kern1pt}

\def\subsubsec#1{\bigskip\noindent{\it{#1}} \bigskip}
\def\del{\partial}

\def\doublefig#1#2#3#4#5{
\par\begingroup\parindent=0pt\leftskip=1cm\rightskip=1cm\parindent=0pt
\baselineskip=11pt \global\advance\figno by 1 \midinsert
\epsfxsize=#4
\centerline{\epsfbox{#2}\hskip0.2in\epsfxsize=#5\epsfbox{#3}} \vskip
12pt {\bf Figure \the\figno:} #1\par
\endinsert\endgroup\par
}
\def\figlabel#1{\xdef#1{\the\figno}}

\def\DDD{B}
\def\CCC{A}

\def\FFF{C}

\def\CC{{\cal C}}

\def\CN{{\cal N}}


\def\ket#1{|#1\rangle}

\def\vev#1{\langle{#1}\rangle}


\noblackbox

\def\IZ{\relax\ifmmode\mathchoice
{\hbox{\cmss Z\kern-.4em Z}}{\hbox{\cmss Z\kern-.4em Z}}
{\lower.9pt\hbox{\cmsss Z\kern-.4em Z}} {\lower1.2pt\hbox{\cmsss
Z\kern-.4em Z}}\else{\cmss Z\kern-.4em Z}\fi}

\font\cmss=cmss10 \font\cmsss=cmss10 at 7pt
\def\IR{\relax{\rm I\kern-.18em R}}
\def\inbar{\,\vrule height1.5ex width.4pt depth0pt}
\def\IC{\relax\hbox{$\inbar\kern-.3em{\rm C}$}}
\def\IR{\relax{\rm I\kern-.18em R}}
\def\IP{\relax{\rm I\kern-.18em P}}
\def\IZ{\relax{\rm Z\kern-.34em Z}}
\def\One{{1\hskip -3pt {\rm l}}}

\def\ie{{\it i.e.}}
\def\frac#1#2{{#1 \over #2}}


\def\journal#1&#2(#3){\unskip, \sl #1\ \bf #2 \rm(19#3) }
\def\andjournal#1&#2(#3){\sl #1~\bf #2 \rm (19#3) }

\def\d{\partial}

%

%
\catcode`\@=11
\def\slash#1{\mathord{\mathpalette\c@ncel{#1}}}
\overfullrule=0pt
\def\AA{{\cal A}}

\def\CC{{\cal C}}
\def\DD{{\cal D}}

\def\HH{{\cal H}}

\def\LL{{\cal L}}

\def\NN{{\cal N}}
\def\OO{{\cal O}}

\def\QQ{{\cal Q}}

\def\underrel#1\over#2{\mathrel{\mathop{\kern\z@#1}\limits_{#2}}}

\def\susy{supersymmetry}

\catcode`\@=12


\def\({\left(}
\def\){\right)}
\def\[{\left[}
\def\]{\right]}
\def\<{\langle}
\def\>{\rangle}
\def\half{{1\over 2}}
\def\d{\partial}
\def\tr{{\rm tr}}
\def\Tr{{\rm Tr}}

\def\det{{\rm det}}
\def\exp{{\rm exp}}

\def\bk{{\bf k}}
\def\bp{{\bf p}}
\def\bx{{\bf x}}

\def\bz{{\bf z}}
\def\bA{{\bf A}}
\def\bV{{\bf V}}
\def\by{{\bf y}}
\def\ba{{\bf a}}
\def\bw{{\bf w}}
\def\bJ{{\bf J}}
\def\bD{{\bf D}}

\def\be{\varepsilon}
\def\bV{{\bf V}}
\def\bP{{\bf P}}
\def\ooint{\relax{\int\kern-.9em ^{\rm O}\kern-.75em _{\rm O}}}
\def\uoint{\relax{\int\kern-.87em ^{\rm O}}}
\def\doint{\relax{\int\kern-.97em _{\rm O}}}
\def\D{\relax{\rm D\kern-0.6em /}}
\def\spi{\relax{\rm \pi\kern-0.5em /}}
\def\sA{\relax{\rm A\kern-0.5em /}}
\def\sp{\relax{\rm p\kern-0.5em /}}

\def\li2{{\rm Li_2}}

\def\ie{{\it i.e.}}
\def\eg{{\it e.g.}}


%


\lref\BuchbinderUB{
  I.~L.~Buchbinder, O.~Lechtenfeld and I.~B.~Samsonov,
  ``N=4 superparticle and super Yang-Mills theory in USp(4) harmonic
  superspace,''
  arXiv:0804.3063 [hep-th].
}

\lref\KovnerVI{
  A.~Kovner and U.~A.~Wiedemann,
  ``Eikonal evolution and gluon radiation,''
  Phys.\ Rev.\  D {\bf 64}, 114002 (2001)
  [arXiv:hep-ph/0106240].
}

\lref\Migdal{
  A.~A.~Migdal,
  ``Loop Equations And 1/N Expansion,''
  Phys.\ Rept.\  {\bf 102}, 199 (1983);
  A.~A.~Migdal,
``Momentum loop dynamics and random surfaces in QCD,"
  Nucl.\ Phys.\  B {\bf 265}, 594 (1986);
  A.~A.~Migdal,
  ``Second quantization of the Wilson loop,''
  Nucl.\ Phys.\ Proc.\ Suppl.\  {\bf 41}, 151 (1995)
  [arXiv:hep-th/9411100].
}

\lref\PolyakovEZ{
  A.~M.~Polyakov,
  {\it Gauge fields and strings},
Chur, Switzerland: Harwood (1987) 301 p.\ (Contemporary
Concepts in Physics, 3).
}

\lref\PolyakovAF{
  A.~M.~Polyakov,
  ``Gauge fields and space-time,''
  Int.\ J.\ Mod.\ Phys.\  A {\bf 17S1}, 119 (2002)
  [arXiv:hep-th/0110196].
}
\lref\PolyakovNC{
  A.~M.~Polyakov,
  ``Confining strings,''
  Nucl.\ Phys.\  B {\bf 486}, 23 (1997)
  [arXiv:hep-th/9607049].
}
\lref\RychkovNI{
  V.~S.~Rychkov,
  ``Wilson loops, D-branes, and reparametrization path-integrals,''
  JHEP {\bf 0212}, 068 (2002)
  [arXiv:hep-th/0204250].
}

\lref\PolyakovJG{
  A.~M.~Polyakov and V.~S.~Rychkov,
  ``Loop dynamics and AdS/CFT correspondence,''
  Nucl.\ Phys.\  B {\bf 594}, 272 (2001)
  [arXiv:hep-th/0005173].
}

\lref\HalpernGD{
  M.~B.~Halpern and P.~Senjanovic,
  ``Functional Bridge Between Gauge Theory And String In Two-Dimensions,''
  Phys.\ Rev.\  D {\bf 15}, 1655 (1977).
}

\lref\HalpernHE{
  M.~B.~Halpern, A.~Jevicki and P.~Senjanovic,
  ``Field Theories In Terms Of Particle-String Variables: Spin, Internal
  Symmetries And Arbitrary Dimension,''
  Phys.\ Rev.\  D {\bf 16}, 2476 (1977).
}

\lref\BrandtGZ{
  R.~A.~Brandt, A.~Gocksch, M.~A.~Sato and F.~Neri,
  ``Loop Space,''
  Phys.\ Rev.\  D {\bf 26}, 3611 (1982).
}

\lref\SorokinNJ{
  D.~P.~Sorokin, V.~I.~Tkach, D.~V.~Volkov and A.~A.~Zheltukhin,
  Phys.\ Lett.\  B {\bf 216}, 302 (1989).
}

\lref\DrummondAQ{
  J.~M.~Drummond, J.~Henn, G.~P.~Korchemsky and E.~Sokatchev,
  ``Hexagon Wilson loop = six-gluon MHV amplitude,''
  arXiv:0803.1466 [hep-th].
}

\lref\BernAP{
  Z.~Bern, L.~J.~Dixon, D.~A.~Kosower, R.~Roiban, M.~Spradlin, C.~Vergu and A.~Volovich,
  ``The Two-Loop Six-Gluon MHV Amplitude in Maximally Supersymmetric Yang-Mills
  Theory,''
  arXiv:0803.1465 [hep-th].
}

\lref\Janik{
  R.~A.~Janik and R.~B.~Peschanski,
  ``High energy scattering and the AdS/CFT correspondence,''
  Nucl.\ Phys.\  B {\bf 565}, 193 (2000)
  [arXiv:hep-th/9907177];
  R.~A.~Janik and R.~B.~Peschanski,
  ``Minimal surfaces and Reggeization in the AdS/CFT correspondence,''
  Nucl.\ Phys.\  B {\bf 586}, 163 (2000)
  [arXiv:hep-th/0003059];
  R.~A.~Janik and R.~B.~Peschanski,
  ``Reggeon exchange from AdS/CFT,''
  Nucl.\ Phys.\  B {\bf 625}, 279 (2002)
  [arXiv:hep-th/0110024].
}

\lref\BuscherSK{
  T.~H.~Buscher,
  ``Path Integral Derivation of Quantum Duality in Nonlinear Sigma Models,''
  Phys.\ Lett.\  B {\bf 201}, 466 (1988).
}

\lref\AldayHR{
  L.~F.~Alday and J.~Maldacena,
  ``Gluon scattering amplitudes at strong coupling,''
  JHEP {\bf 0706}, 064 (2007)
  [arXiv:0705.0303 [hep-th]].
}

\lref\PolyakovJU{
  A.~M.~Polyakov,
``The wall of the cave,''
  Int.\ J.\ Mod.\ Phys.\  A {\bf 14}, 645 (1999)
  [arXiv:hep-th/9809057].
}

\lref\DrummondAU{
  J.~M.~Drummond, G.~P.~Korchemsky and E.~Sokatchev,
  ``Conformal properties of four-gluon planar amplitudes and Wilson loops,''
  arXiv:0707.0243 [hep-th].
}

\lref\BrandhuberYX{
  A.~Brandhuber, P.~Heslop and G.~Travaglini,
  ``MHV Amplitudes in N=4 Super Yang-Mills and Wilson Loops,''
  arXiv:0707.1153 [hep-th].
}

\lref\Schubert{
  C.~Schubert,
 ``Perturbative quantum field theory in the string-inspired formalism,''
  Phys.\ Rept.\  {\bf 355}, 73 (2001)
  [arXiv:hep-th/0101036].
}

\lref\StrasslerZR{
  M.~J.~Strassler,
  ``Field theory without Feynman diagrams: One loop effective actions,''
  Nucl.\ Phys.\  B {\bf 385}, 145 (1992)
  [arXiv:hep-ph/9205205].
}

\lref\KorchemskayaQP{
  I.~A.~Korchemskaya and G.~P.~Korchemsky,
  ``High-energy scattering in QCD and cross singularities of Wilson loops,''
  Nucl.\ Phys.\  B {\bf 437}, 127 (1995)
  [arXiv:hep-ph/9409446].
}

\lref\BrandtGZ{
  R.~A.~Brandt, A.~Gocksch, M.~A.~Sato and F.~Neri,
  ``Loop Space,''
  Phys.\ Rev.\  D {\bf 26}, 3611 (1982).
}
\lref\Srednicki{
  M.~Srednicki,
  ``Quantum field theory,''
{\it  Cambridge,
UK: Univ. Pr. (2007) 641 p}
}

\lref\BrinkUF{
  L.~Brink, P.~Di Vecchia and P.~S.~Howe,
  ``A Lagrangian Formulation Of The Classical And Quantum Dynamics Of Spinning
  Particles,''
  Nucl.\ Phys.\  B {\bf 118}, 76 (1977).
}

\lref\AndreevBZ{
  O.~D.~Andreev and A.~A.~Tseytlin,
``Generating functional for scattering amplitudes
and effective action in the open superstring theory,"
  Phys.\ Lett.\  B {\bf 207}, 157 (1988).
}

\lref\MaldacenaIM{
  J.~M.~Maldacena,
 ``Wilson loops in large N field theories,''
  Phys.\ Rev.\ Lett.\  {\bf 80}, 4859 (1998)
  [arXiv:hep-th/9803002].
}

\lref\ReyIK{
  S.~J.~Rey and J.~T.~Yee,
  ``Macroscopic strings as heavy quarks in large N gauge theory and  anti-de
  Sitter supergravity,''
  Eur.\ Phys.\ J.\  C {\bf 22}, 379 (2001)
  [arXiv:hep-th/9803001].
}

\lref\BalachandranYA{
  A.~P.~Balachandran, P.~Salomonson, B.~S.~Skagerstam and J.~O.~Winnberg,
  Phys.\ Rev.\  D {\bf 15}, 2308 (1977);
  J.~Ishida and A.~Hosoya,
  Prog.\ Theor.\ Phys.\  {\bf 62}, 544 (1979); J.~L.~Gervais and A.~Neveu,
  Nucl.\ Phys.\  B {\bf 163}, 189 (1980); A.~Barducci, R.~Casalbuoni and L.~Lusanna,
  Nucl.\ Phys.\  B {\bf 180}, 141 (1981);
  D.~Friedan and P.~Windey,
  Nucl.\ Phys.\  B {\bf 235}, 395 (1984).
}

\lref\HellermanBU{
  S.~Hellerman, S.~Kachru, A.~E.~Lawrence and J.~McGreevy,
  ``Linear sigma models for open strings,''
  JHEP {\bf 0207}, 002 (2002)
  [arXiv:hep-th/0109069].
}

\lref\SchwingerXK{
  J.~S.~Schwinger,
  Phys.\ Rev.\  {\bf 82}, 914 (1951); Z.~Bern and D.~A.~Kosower,
  Nucl.\ Phys.\  B {\bf 362}, 389 (1991).
}

\lref\MarcusCM{
  N.~Marcus and A.~Sagnotti,
  ``Group Theory From Quarks At The Ends Of Strings,''
  Phys.\ Lett.\  B {\bf 188}, 58 (1987).
}

\lref\FriedanXR{
  D.~Friedan and P.~Windey,
  ``Supersymmetric Derivation Of The Atiyah-Singer Index And The Chiral
  Anomaly,''
  Nucl.\ Phys.\  B {\bf 235}, 395 (1984).
}

\lref\McGreevyKT{
  J.~McGreevy and A.~Sever,
  ``Quark scattering amplitudes at strong coupling,''
  arXiv:0710.0393 [hep-th].
}

\lref\PolyakovTI{
  A.~M.~Polyakov and V.~S.~Rychkov,
 ``Gauge fields - strings duality and the loop equation,''
  Nucl.\ Phys.\  B {\bf 581}, 116 (2000)
  [arXiv:hep-th/0002106].
}

\lref\MathurTP{
  S.~D.~Mathur,
  ``Is the Polyakov path integral prescription too restrictive?,''
  arXiv:hep-th/9306090.
}

\lref\BernEW{
  Z.~Bern, M.~Czakon, L.~J.~Dixon, D.~A.~Kosower and V.~A.~Smirnov,
  Phys.\ Rev.\  D {\bf 75}, 085010 (2007)
  [arXiv:hep-th/0610248]; Z.~Bern, J.~J.~M.~Carrasco, H.~Johansson and D.~A.~Kosower,
  Phys.\ Rev.\  D {\bf 76}, 125020 (2007)
  [arXiv:0705.1864 [hep-th]];
  J.~M.~Drummond, J.~Henn, V.~A.~Smirnov and E.~Sokatchev,
  JHEP {\bf 0701}, 064 (2007)
  [arXiv:hep-th/0607160]; J.~M.~Drummond, G.~P.~Korchemsky and E.~Sokatchev,
  arXiv:0707.0243 [hep-th].
}

\lref\AharonyRQ{
  O.~Aharony and Z.~Komargodski,
  JHEP {\bf 0801}, 064 (2008)
  [arXiv:0711.1174 [hep-th]].
}

\lref\WittenQS{
  E.~Witten,
  Nucl.\ Phys.\  B {\bf 276}, 291 (1986).
}

\lref\AldayHE{
  L.~F.~Alday and J.~Maldacena,
  ``Comments on gluon scattering amplitudes via AdS/CFT,''
  JHEP {\bf 0711}, 068 (2007)
  [arXiv:0710.1060 [hep-th]].
}

\lref\AldayMF{
  L.~F.~Alday and J.~M.~Maldacena,
  ``Comments on operators with large spin,''
  JHEP {\bf 0711}, 019 (2007)
  [arXiv:0708.0672 [hep-th]].
}

\lref\PolyakovTJ{
  A.~M.~Polyakov,
  ``String theory and quark confinement,''
  Nucl.\ Phys.\ Proc.\ Suppl.\  {\bf 68}, 1 (1998)
  [arXiv:hep-th/9711002].
}

\lref\BuchbinderUB{
  I.~L.~Buchbinder, O.~Lechtenfeld and I.~B.~Samsonov,
  arXiv:0804.3063 [hep-th].
}

\lref\DrukkerZQ{
  N.~Drukker, D.~J.~Gross and H.~Ooguri,
  ``Wilson loops and minimal surfaces,''
  Phys.\ Rev.\  D {\bf 60}, 125006 (1999)
  [arXiv:hep-th/9904191].
}

\lref\MigdalRN{
  A.~A.~Migdal,
  ``Hidden symmetries of large N {QCD},''
  Prog.\ Theor.\ Phys.\ Suppl.\  {\bf 131}, 269 (1998)
  [arXiv:hep-th/9610126].
}

\lref\eikonalrefs{
  H.~D.~I.~Abarbanel and C.~Itzykson,
  ``Relativistic eikonal expansion,''
  Phys.\ Rev.\ Lett.\  {\bf 23}, 53 (1969);
    R.~L.~Sugar and R.~Blankenbecler,
  Phys.\ Rev.\  {\bf 183}, 1387 (1969).
    A.~I.~Karanikas and C.~N.~Ktorides,
  ``Worldline approach to forward and fixed angle fermion fermion  scattering
  in Yang-Mills theories at high energies,''
  Phys.\ Rev.\  D {\bf 59}, 016003 (1999)
  [arXiv:hep-ph/9807385].
}

\lref\heavyquark{
A. V. Manohar and M. B. Wise, {\it Camb.\ Monogr.\ Part.\ Phys.\ Nucl.\ Phys.\ Cosmol.\ } {\bf 10}, 1 (2000);
M. Neubert,  Phys.\ Rept.\ {\bf 245}, 259 (1994), hep-ph/9306320;
  G.~P.~Korchemsky and A.~V.~Radyushkin,
``Infrared factorization, Wilson lines and the heavy quark limit,''
  Phys.\ Lett.\  B {\bf 279}, 359 (1992)
  [arXiv:hep-ph/9203222].
}

\lref\EichtenKD{
  E.~Eichten and R.~Jackiw,
  ``Failure of the eikonal approximation for the vertex function in a boson
  field theory,''
  Phys.\ Rev.\  D {\bf 4}, 439 (1971).
}

\lref\LRW{
  H.~Liu, K.~Rajagopal and U.~A.~Wiedemann,
  ``Calculating the jet quenching parameter from AdS/CFT,''
  Phys.\ Rev.\ Lett.\  {\bf 97}, 182301 (2006)
  [arXiv:hep-ph/0605178].
}

\lref\BrezinSK{
  E.~Brezin and J.~Zinn-Justin,
  {\it Fields, Strings And Critical Phenomena. Proceedings, 49th Session Of The
  Les Houches Summer School In Theoretical Physics, Nato Advanced Study
  Institute, Les Houches, France, June 28 - August 5, 1988,}
Amsterdam, Netherlands: North-Holland (1990) 640 p. }

\lref\KaranikasHZ{
  A.~I.~Karanikas and C.~N.~Ktorides,
  ``Extension of worldline computational algorithms for {QCD} to open
  fermionic contours,''
  JHEP {\bf 9911}, 033 (1999)
  [arXiv:hep-th/9905027].
}

\lref\AvramisXF{
  S.~D.~Avramis, A.~I.~Karanikas and C.~N.~Ktorides,
  ``Perturbative computation of the gluonic effective action via  Polyakov's
  world line path integral,''
  Phys.\ Rev.\  D {\bf 66}, 045017 (2002)
  [arXiv:hep-th/0205272].
}

\Title{\vbox{\baselineskip12pt \hbox{arXiv:0806.0668} \hbox{BRX
TH-597}\hbox{MIT-CTP/3952} }} {\vbox{ \centerline{Planar scattering
amplitudes from Wilson loops}
\smallskip
\smallskip
\smallskip
}} \centerline{John McGreevy${}^1$ and Amit Sever${}^2$}

\bigskip

\centerline{{$^1$Center for Theoretical Physics, Massachussetts
Institute of Technology}} \centerline{{Cambridge, MA 02139, USA}}
\centerline{{$^2$Brandeis Theory Group, Martin Fisher School of
Physics, Brandeis University}} \centerline{{Waltham, MA 02454, USA}}

\bigskip
\bigskip
\bigskip
\noindent

We derive an
expression for parton scattering amplitudes of planar gauge theory
in terms of sums of Wilson loops. We study in detail the example of
Yang-Mills theory with an adjoint Higgs field. The expression
exhibits the T-duality performed by Alday and Maldacena in the AdS
dual as a Fourier transform in loop space. When combined with the
AdS/CFT correspondence for Wilson loops and a strong coupling
argument for the dominance of 1PI diagrams, this leads to a
derivation of the Alday-Maldacena holographic prescription for
scattering amplitudes in terms of momentum Wilson loops. The formula
leads to a conjecture for a relationship between position-space and
momentum-space Wilson loops in $\NN=4$ SYM at finite coupling.

\Date{June, 2008}


\newsec{Introduction}

Imagine that you are so powerful that given an arbitrary gauge
theory, you can calculate the expectation value of the Wilson loop
around an arbitrary path, as a functional of the path. This would be
a lot of information about the gauge theory. In particular, it
should determine the scattering amplitudes of the partons of the
theory.
How would one extract this information?

In this paper we construct a prescription which extracts the parton
scattering amplitudes from Wilson loop expectation values.
The result simplifies dramatically in the 't Hooft limit,
and we focus on this case.
A formula similar to our result was conjectured by
Polyakov, based on string theory intuition \PolyakovJU. That formula
(equation (2.5) below), literally applies to interactions mediated
by a scalar field. Much of our effort in this paper will be devoted
to generalizing this for interactions mediated by a gauge field.

The immediate motivation for our work was the connection between
parton scattering amplitudes and Wilson loops exposed by the work of
Alday and Maldacena (AM) \refs{\AldayHR, \AldayHE, \AldayMF}.
Our results give a partial derivation of the Alday-Maldacena
prescription in terms of previously-understood entries in the
AdS/CFT dictionary.

Our work
may shed light on the mysterious relation between
position space loops and momentum space loops suggested by the work
of \AldayHR. Specifically, matching our prescription to the
correctness of the AM result for gluon scattering at large
$\lambda$, combined with the weak coupling relation between gluon
scattering and position space Wilson loops, suggests a conjecture
for the precise relationship, elaborated in section 5.

It may also teach us something about how a string theory emerges
from the gauge theory. One of Polyakov's stated motivations for
studying the formula (2.5) was to understand the action of the loop
operator in the string language \PolyakovTJ. We comment on this at
the end.

The use of Wilson loops in the study of scattering has a long
history.  Attempts to reformulate gauge theory in terms of loop space
\refs{\HalpernGD, \HalpernHE, \BrandtGZ, \Migdal, \MigdalRN,
\DrukkerZQ, \PolyakovJG} have occasionally resulted in related
formulae, none of which precisely met our needs.
We were unable to find
in the literature an answer to the question
posed in the first paragraph, namely an expression for the scattering
amplitude as a sum over Wilson loops.
Previous attempts to relate scattering amplitudes to Wilson
loops in string theory using the eikonal approximation include
\Janik.

The paper is organized as follows. In the next section we begin our
study of planar gauge theory scattering amplitudes and make a first
pass at a worldline description. Then we show that the
mysterious-seeming T-duality used by Alday and Maldacena in finding
their saddle point has a simple interpretation in terms of a Fourier
transform in loop space \PolyakovEZ. In section four, we give a
systematic treatment. In section five, we discuss the most important
difference between the heuristic (2.5) and the correct formula,
namely, the fact that the Wilson sums give only the 1PI effective
action, which must be connected in trees to get the full scattering
amplitude. In section six we work out the prescription in detail in
an example. After that we discuss some possible further applications
of the prescription. Sequestered to the three appendices are our
discussions of reparametrization-invariant worldline theories
(\CCC), worldline superspace (\DDD), and the action of the loop
operator (\FFF).

\newsec{Gluon scattering amplitudes}

A well known question in QED is the following: The electron field
operator is not gauge invariant. However, there are physical states
in the theory where an electron is localized around some point
$\bx$. So, what is an operator that creates from the vacuum a state
with an electron at the point $\bx$? An example of such an operator is
the electron field operator attached to a Wilson line from the point
$\bx$ to infinity. This operator is charged only under the global
part of the gauge group, which is not gauged; this global charge is
the electron charge. Given such an operator, one can then consider
its Green functions and obtain scattering amplitudes from them via
an LSZ formula. A consistency requirement of such manifestly gauge
invariant definition of electron scattering amplitudes is that it
will reproduce the known perturbative expansion of the amplitude.

In this paper, we will derive a gauge invariant expression for {\it
planar} parton scattering amplitudes in non-abelian gauge theory.
Similarly to electrons in QED, these are charged only under the
global part of the gauge group. Although the basic block in our
planar expression will be the Wilson loop, it also can be expanded
as
a reorganization of planar Feynman
graphs and therefore it will automatically reproduce the perturbative
expansion. In this subsection we start with an intuitive guess for
such an expression.
This cartoon of the formula
does
not reproduce the known perturbative expansion of the amplitude.
However, it captures a lot of the physics of the problem and will be
presented first to develop intuition.

A gluon is characterized by its momentum $\bk_i$, its polarization
vector $\varepsilon_i$ and an adjoint color matrix $T^{a_i}$. In
perturbation theory, a gluon scattering amplitude can be
rearranged as a sum over all possible color contractions of the
external gluons. The coefficients of the color traces are
independent of the specific color matrices in the trace and are
called {\it partial amplitudes}. In the planar limit, only single
traces contribute and the amplitude takes the form
\eqn\gluonsA{\AA_n^{planar}=\sum_\pi\tr\(T^{\pi(1)}\dots
T^{\pi(n)}\)
A_n((\bk_{\pi(1)},\varepsilon_i),\dots,(\bk_{\pi(n)},\varepsilon_n))~.}
In this paper we will be interested only in these color-ordered partial
amplitudes.
%

\subsec{A first pass at scattering amplitudes from Wilson loops}

Based on string theory intuition, Polyakov \PolyakovJU\ suggested
the following expression for the gluon partial amplitude in QCD
\eqn\ampWilson{A_n^P\equiv\int[D\bx]\prod_i\int_{s_{i-1}} ds_i\,
\varepsilon_i^\mu{dx_\mu\over ds}e^{ i \bk_i \cdot
\bx(s_i)}\<W[\bx(\cdot)]\>~,}
where
\eqn\polint{\int[D\bx] \dots \equiv \int_0^\infty{dT\over
T}\NN\int\[D\bx\]_1 e^{-\half\int_0^T\dot\bx^2 ds}\dots}
is the integral over all closed loops with measured with respect to
the induced metric on the worldline \PolyakovEZ. The integral
$\int[D\bx]_1$ in \polint\ is an integral over all closed curves
defined with respect to a trivial worldline metric. That is, it is
defined by dividing the segment $[0,T]$ into infinitesimal pieces
which are equally-spaced in the parameter $s$:
\eqn\dxone{\int[D\bx]_1 \equiv \prod_{\ell} d^Dx(s_\ell)~,}
where $D=4$ is the number of spacetime dimensions. The normalization
constant $\NN$ is defined such that in the continuum limit
$$\NN\int[D\bx(\cdot)]_1e^{-\half\int_0^T\dot\bx^2 ds}=[2\pi
T]^{-D/2}~.$$
As in open string theory, the terms
$$\varepsilon_i^\mu{dx_\mu\over ds}e^{
i \bk_i \cdot \bx(s_i)}$$
are gluon like vertex insertion ordered and integrated along the
loop. Each closed loop is weighted by the expectation value of
corresponding Wilson loop $\<W[x(\cdot)]\>$.

From the open string point of view, the loop represent the boundary
of an open string and the Wilson loop expectation value represent
the worldsheet path of an open string (embedded in one higher
dimension) with a fixed boundary loop. So equation \ampWilson\
stands somewhere between the gauge theory and the string theory
descriptions where we separate the dynamics of the boundary of the
open string from the dynamics of its bulk, represented in the gauge
theory languish by the Wilson loop expectation value.

The generalization of \ampWilson\ for finite worldline mass $m$ is
\eqn\mass{A_n^\bP=\int{dT\over
T}\NN\int[D\bx(\cdot)]_1e^{-\int_0^Tds\(\half\dot\bx^2+m^2\)}F[\bx(\cdot);\{\varepsilon_i,\bk_i\}]~,}
where
$$F[\bx(\cdot);\{\varepsilon_i,\bk_i\}]
=\prod_i\int_{s_{i-1}}^T ds_i\, \varepsilon_i\cdot\dot\bx(s_i)e^{ i
\bk_i \cdot \bx(s_i)}\<W[\bx(\cdot)]\>~.$$
%
$F[\bx(\cdot);\{\varepsilon_i,\bk_i\}]$
is reparametrization invariant. Therefore, $A_n^\bP$ can also be
written in terms of a Nambu-Goto-like action \PolyakovEZ
\eqn\Polyakovrep{\eqalign{A_n^\bP=&\int{[D\bx(\cdot)]_{\dot\bx^2}\over
[Df]} e^{-m_0\int_0^1\sqrt{\dot\bx^2}}
F[\bx(\cdot);\{\varepsilon_i,\bk_i\}]\cr =&
\int{[D\bx(\cdot)]_{\dot\bx^2}\over [Df]}
e^{-m_0\int_0^1\sqrt{\dot\bx^2}} \prod_i\int_{s_{i-1}}^1 ds_i\,
\varepsilon_i\cdot\dot\bx(s_i)e^{ i \bk_i \cdot
\bx(s_i)}\<W[\bx(\cdot)]\>~,}}
where $\int[D\bx(\cdot)]_{\dot\bx^2}$ is an integral over all closed
curves with respect to the induced metric on the worldline, $[Df]$
stands for the volume of the gauge (reparametrization) group and
$m_0$ is the bare mass which is dialed such that \Polyakovrep\
admits a continuum limit with a physical mass $m$.

We can interpret $A_n^P$ as follows: Inserting a
spacelike Wilson loop, even a smooth one, into a Lorentzian gauge
theory is a huge perturbation of the vacuum.
Since the loop is a color source with short-distance structure,
there is a large
amplitude to create many very hard gluons, and hence a very small
amplitude not to do so. In the formula for the scattering amplitude
in terms of Wilson loops, the integral over contours constructs a
superposition of loops which create a fixed number of external gauge
bosons.

The expression for the gluon amplitude in terms of closed Wilson
loops \ampWilson\ has a natural generalization for scattering of
quarks.
In this paper however, we will
discuss the scattering of adjoint fields only.

Equation \ampWilson\ is roughly of the form that we will derive at
large $N$ using the worldline description of one loop determinants.
We label it with a \bP\ to distinguish it from the correct formula
which we will eventually write. As we will see, although the formal
expression \ampWilson\ captures much of the physics of the
amplitude, it is not correct for several reasons. First, even in
pure YM theory, the Wilson loop is the phase acquired by a scalar
field, not a vector field. Accounting correctly for the spin
modifies both the ``gluon vertex operator'' and the form of the
Wilson loop. Second, the amplitude has IR divergences and requires
regularization. As we will discuss, one way to regularize the
amplitude is by turning on a worldline mass \mass. Thirdly, an
equation in the spirit of \ampWilson\ holds only in the planar limit
and in general gets (tractable) $1/N$ corrections. Finally, we will
see that even at large $N$ the correct formula is a sum over
tree-level Feynman diagrams using objects like \ampWilson\ as
vertices.

To see that some correction to $A^{\bP}$ will be needed,
even perturbatively,
note that to lowest order in the 't Hooft coupling $\lambda$, it
gives only a scalar one loop contribution to the amplitude
\StrasslerZR, while any tree level contribution is absent.
Non-perturbatively, we can see the need for improvement
as follows.
The amplitude $A_n$ is invariant under shift of a
polarization vector by the corresponding momenta, (as is necessary
for gauge invariance)
\eqn\Gaugeinv{A_n\[\dots,(\bk_i,\be_i+c\,\bk_i),\dots\]=
A_n\[\dots,(\bk_i,\be_i),\dots\]~,}
where $c$ is a constant parameter. In $A^\bP$ \ampWilson, we note that
a longitudinally-polarized gluon vertex
$$\bk\cdot{d\bx\over
ds}\,e^{i\bk\cdot\bx(s)}=-i{d \over ds}\,e^{i\bk\cdot\bx(s)}$$
is a total derivative. However, in the partial amplitude, $s_i$ is
integrated only on the segment $\[s_{i-1},s_{i+1}\]$. Therefore,
instead of \Gaugeinv\ we find
\eqn\noncancelled{
\eqalign{&A^{\bP}_n\[\dots,(\bk_i,\be_i+c\,\bk_i),\dots\]-
A^{\bP}_n\[\dots,(\bk_i,\be_i),\dots\]\cr =&\,
icA^{\bP}_{n-1}\[\dots,(\bk_{i-1}+\bk_i,\be_{i-1}),(\bk_{i+1},\be_{i+1}),\dots\]\cr
-&icA^{\bP}_{n-1}\[\dots,(\bk_{i-1},\be_{i-1}),(\bk_{i+1}+\bk_i,\be_{i+1}),\dots\]~.}}
If $\bk_i$ and $\bk_{i+1}$ are not colinear, then $\bk_i+\bk_{i+1}$
is not null. As will become clear, the expression $A^\bP$ does not
vanish off-shell. Here we note that even if it would have been zero
for off-shell gluon momenta, the right hand side of \noncancelled\
would not be zero in the case were two adjacent gluons are colinear.
In string theory such corrections are absent, a conclusion which
follows from the {\it canceled propagator argument}. Analyticity
implies that the amplitude
with a longitudinal external state must vanish for all values of the
$\bk$'s.
We will see below that the correct loop-sum formula for the
scattering amplitude will have additional contributions
which can restore gauge
invariance \Gaugeinv\ of the amplitude.

Note that for $n=2$, the two color orderings of the external gluons
are the same. In that case $s_1$ and $s_2$ can be independently
integrated over the whole loop, and therefore $A_2^\bP=0$ whenever a
polarization vector is longitudinal. The integral over the $\bx$
zero-mode leads to a momentum conservation delta function. We
therefore conclude that
\eqn\scalarmass{A_2^\bP[(\be_1,\bk_1),(\be_2,\bk_2)]\propto
\delta(\bk_1+\bk_2)~\be_1^\mu\be_2^\nu\(k_\mu
k_\nu-\bk^2\eta_{\mu\nu}\)G(\bk^2)~,}
where $G$ is some smooth function.

In the \S4, we will {\it derive} the correct analog of
\ampWilson\ for any planar gauge theory in a specific regularization
inspired by the Alday-Maldacena $z_{IR}$ regularization. We will
interpret \scalarmass\ as the scalar correction to the planar gluon
propagator. Along the way we will see where and how $1/N$
corrections can be incorporated.

\newsec{T duality and loop space Fourier transform}

Let $\CC_\bx$ represent a closed loop and let $F[\CC_x]$ be some
functional on loop space. That is, if $\bx(s)$ is some
parametrization of $\CC_\bx$, then $F[\bx(\cdot)]$ depends only on
the image of $\bx(s)$ and not on its specific parametrization. An
example of such
a functional is the expectation value of a Wilson loop
operator in $SU(N)$ gauge theory:
\eqn\Wilson{W\[\CC_\bx\]={1\over N}\<\tr Pe^{\oint\bA\cdot
d\bx}\>~,}
where $\oint\bA\cdot d\bx=\int\bA(\bx(s))\cdot\dot\bx(s)ds$.

Let $\CC_\bp$ be some closed loop in momentum space and let
$\bp(s),~s\in[0,1]$ be some parametrization of $\CC_\bp$. We define
a Fourier transform of $F[\CC_\bx]$ along $\bp(\cdot)$ to be
\eqn\lsft{\widetilde
F[\bp(\cdot)]\equiv\int[\DD\bx]e^{i\oint\bp\cdot d\bx}F[\bx(\cdot)]}
and the corresponding {\it momentum Wilson loop} to be
\eqn\MomentumWilson{\<\widetilde
W[\bp(\cdot)]\>\equiv\int[\DD\bx]e^{i\oint\bp\cdot
d\bx}\<W[\bx(\cdot)]\>~,}
where $\int[\DD\bx]$ is an integral over all closed curves. In
\lsft\ we have not specified the measure with respect to which the
integral over closed curves is taken.
Different such measures lead to
different Fourier transforms,
and in general the result depends of the
specific parametrization of $\CC_\bp$, $\bp(\cdot)$. Here we would
like to mention one such measure considered by Migdal \Migdal. This
is the measure where the worldline metric in $\int[\DD\bx]$ is the
induced metric from the map $\bp(s)$ to momentum space. This measure
is special
because the result depends only on $\CC_\bp$ and
not on its specific parametrization.\foot{Of course replacing the
measure $\int[\DD\bx]_{Migdal}$ by $\int[\DD\bx]_{Migdal} G[\CC_x]$
for some reparametrization-invariant functional $G$ will lead to
another reparametrization-invariant Fourier transform.}
Nevertheless, this is not the measure we will derive
from
the scattering amplitude;
what we will find is closer to what we would write by analogy with
string theory.

Note that for any point in the integral over the gluon insertion
points $\{s_i\}$, the momentum dependence of $A_n^\bP$ can be
written as
$$\sum\bk_i\cdot\bx(s_i)=\int ds
\sum\bk_i\cdot\bx(s)\delta(s-s_i)=-\int_0^T ds~
\bp\({s\over T}\)\cdot\dot\bx(s)~,$$
where $\bp(s)$ is the polygon momentum loop
\eqn\momentuml{\bp(s)=\sum_i \bk_i\theta(s-s_i/T)~.}
The heuristic scattering formula \ampWilson\ can
therefore be written
in terms of the momentum loop \momentuml:
\eqn\dressing{A_n^\bP=Z_3^{-n}\prod_i\int_{s_{i-1}}^1
ds_i\varepsilon_i^\mu{\delta\over\delta p^\mu(s_i)}\<\widetilde
W[\bp(\cdot)]\>~,}
where
$$\<\widetilde W[\bp(\cdot)]\>=\int_0^\infty{dT\over T}\NN\int [D\bx]_1
~e^{-\half\int_0^T\dot\bx^2 ds} \<W[\bx(\cdot)]\>e^{-i\oint\bp\cdot
d\bx}$$
is a momentum Wilson loop, defined with respect to the measure for
loops discussed in the previous section. This Fourier transform is,
however, not reparametrization invariant. A reparametrization
invariant result is obtained only after dressing $\<\widetilde
W[\bp(\cdot)]\>$ into $A_n^\bP$ \dressing. This dressing \dressing\
can be thought of as an integration over all reparametrizations of
the polygon $\bp(s)$ \momentuml.

One of the properties of momentum-space Wilson loops is that
$\bp(s)$ can have discontinuities. In the study of gluon scattering
amplitudes, such discontinuities naturally appear as the T-dual of
the gluon momenta.

In \AldayHR\ Alday and Maldacena gave a prescription for the
holographic computation of gluon scattering amplitudes using open
strings in AdS. The prescription amounts to
computing an open string
amplitude on a probe brane in AdS. At large 't Hooft coupling, the
open string amplitude is approximated by its saddle point.
In that
approximation, one can neglect the polarization dependence of the
scattered gluons. The open string amplitude is expressed through an
integral over the moduli space of the
insertion points of the gluon vertex
operators. In order to find the saddle point, Alday
and Maldacena first did a T-duality along the 3+1 transverse
directions (see \McGreevyKT\ for details). That T-duality commutes
with the integral over the insertion points and can be
done point by point.

Next, we show that after suppressing the polarization dependence of
the gluon vertex insertions, this T-duality
amounts to a Fourier transform in loop space. More generally,
T-duality takes the form of a Fourier transform in loop space
whenever one studies an open string that ends on some closed curve
in a specific parametrization. A meaningful result may then be
obtained by summing over all possible reparametrizations of the curve
or by a saddle point approximation to that sum \AldayHR.

To see this, consider the string dual of the Wilson loop \Wilson.
That is, consider an open string in $AdS_5$ with boundary conditions
such that it ends on the curve $\bx_b(s)$ at the boundary of AdS (at
$z_{UV}\to 0$). In Poincar\'e AdS
\eqn\Poincare{ds^2=R_{AdS}^2{dz^2+dx_{3+1}^2\over z^2}~,}
the string worldsheet action in conformal gauge is
\eqn\xWSact{S_0={\sqrt\lambda\over 4\pi}\int_\DD d\sigma\d\tau\
\[(\d_\alpha\bx)^2+(\d_\alpha z)^2\]/z^2~.}
The Wilson loop expectation value is roughly (ignoring the ghosts
and other worldsheet fields for the moment)
\eqn\Wexpect{\<W\[\CC_\bx\]\>=\int\[Df(\sigma)\]\int
\[D\bx(\sigma,\tau)\]_{\bx(\sigma,0)=\bx_b(f(\sigma))}
D\[z(\sigma,\tau)\]_{z(\sigma,0)=0}e^{-S_0}~,}
where the $f$-integral is over the group of boundary
reparametrizations \RychkovNI, and $\bx_b(\sigma)$ is some
non-degenerate parametrization of $\CC_\bx$. Next, we do a change of
variables in the path integral which can be described as a
``T-duality'' along the non-compact 3+1 flat directions. To do this,
we follow Buscher \BuscherSK. For each field $x^\mu$, we gauge the
shift symmetry $ x^\mu \to x^\mu + \lambda^\mu$, and introduce a
worldsheet gauge field ${\it V}_\alpha^\mu$ and a scalar lagrange
multiplier $p^\mu$. We then consider the gauge-invariant action
\eqn\Bucher{S_1={\sqrt\lambda\over 4\pi}\int_\DD d\sigma\d\tau\
\[(\d_\alpha
\bx-\bV_\alpha)^2/z^2-i\bp \cdot{\bf F}\]~,}
where ${\bf F}=\d_\tau\bV_\sigma-\d_\sigma\bV_\tau$ and we are
suppressing the kinetic term for $z$. Next, we fix a gauge by
absorbing $d\bx$ into the gauge field. The resulting gauge field
$\bV$ is subject to the boundary condition
$\bV_\sigma(\sigma,0)=\d_\sigma\bx_b(\sigma)$. To see this,
introduce a boundary auxiliary field ${\bf b}(\sigma)$ and rewrite
the Dirichlet boundary conditions for $\bx$ by adding to the
boundary action the term
\eqn\xbc{i\int d\sigma{\bf
b}(\sigma)\cdot\[\bx(\sigma,0)-\bx_b(\sigma)\]~.}
Now when we gauge the shift symmetry of $\bx$, \xbc\ is replaced by
\eqn\xbcg{S_2=i\int d\sigma{\bf
b}(\sigma)\cdot\[[\bx(\sigma,0)-\bx(0,0)]-[\int_0^\sigma
ds\bV_\sigma(s)-\bx(0,0)]-\bx_b(\sigma)\]~.}
After we absorb $d\bx$ into the gauge field, the term
$[\bx(\sigma,0)-\bx(0,0)]$ is removed from \xbcg. Integrating over
${\bf b}(\sigma)$ yields
\eqn\Abc{\delta\(V_\sigma-\d_\sigma\bx_b\)\delta\(\bx(0,0)-\bx_b(0)\)~.}
If we first integrate out $\bp$, then $\bV$ becomes a flat
connection $V=d\tilde\bx$. By defining
$\tilde\bx(0,0)=\bx(0,0)=\bx_b(0)$, we see that \Bucher\ is
equivalent to the original action. If on the other hand, we first
integrate $\bV$, then it is convenient to integrate by parts in the
second term in \Bucher. We then have
\eqn\Buchertwo{\eqalign{S_1+S_2=&{\sqrt\lambda\over 4\pi}\int_\DD
d\sigma\d\tau\
\[(\bV_\alpha\cdot\bV_\alpha)/z^2+i\(\bV_\sigma\cdot\d_\tau\bp
-\bV_\tau\cdot\d_\sigma\bp \)\]-i{\sqrt\lambda\over
4\pi}\int_{\d\DD} d\sigma {\bf V}_\sigma\cdot\bp \cr -&i\int_{\d D}
d\sigma{\bf b}(\sigma)\cdot\[\int_0^\sigma
ds\bV_\sigma(s)-\bx(0,0)+\bx_b(\sigma)\]~.}}
Now, by integrating out $\bV$ we get
\eqn\Aeom{\bV_\sigma=-iz^2\d_\tau\bp~,\qquad\bV_\tau=iz^2\d_\sigma\bp~,\qquad
{\bf b}(\sigma)={\sqrt\lambda\over 4\pi}\d_\sigma\bp~.}
By plugging \Aeom\ back into \Buchertwo\ the action for $\bp$
becomes
\eqn\yWSact{\eqalign{S_3=&{\sqrt\lambda\over 4\pi}\int_\DD
d\sigma\d\tau\
\[z^2(\d_\alpha\bp)^2+(\d_\alpha z)^2/z^2\]-i{\sqrt\lambda\over
4\pi}\int_{\d\DD} d\sigma \d_\sigma\bp\cdot[\bx_b-\bx(0,0)]\cr
=&{\sqrt\lambda\over 4\pi}\int_\DD d\sigma\d\tau\
\[z^2(\d_\alpha\bp)^2+(\d_\alpha z)^2/z^2\]-i{\sqrt\lambda\over
4\pi}\int_{\d\DD} d\sigma \d_\sigma\bp\cdot\bx_b~.}}
If we now rescale $\bp$ and $z$ as
$\(\bp,z\)\to\(-{4\pi\over\sqrt\lambda}\bp,{\sqrt\lambda\over 4\pi}
z\)$, then \MomentumWilson\ is reproduced if we identify the T-dual
Wilson loop with an open string in dual AdS that ends on the
Poincar\'e horizon. The dual AdS metric is
\eqn\DPoincare{ds^2=R_{AdS}^2{dr^2+dp_{3+1}^2\over r^2}~,}
where the radial direction is $r={1\over z}$.\foot{Note that if we
change variables in the path integral from $z$ to $r={1\over z}$,
then we generate a dilaton $\Phi\sim\log(r)$.} The path integral
over $\bp(\sigma,\tau)$ splits into an integral over its boundary
value $\bp(\sigma,0)=\bp_b(\sigma)$ and an integral over its bulk
value $\bp(\sigma,\tau>0)$, subject to the Dirichlet
 boundary condition, $\bp(\sigma, \tau = 0 ) = \bp_b(\sigma)$.

To summarize, we have shown that
\eqn\summary{\eqalign{&\int \[Dz\]_{z_\d=0}\[D\bx\]_{\bx_\d=\bx_b}
e^{-S_0[\bx,z]}=\int\[Dz\]_{z_\d=0}\[D\bx\]\[D{\bf b}\]
\[D\bV\]\[D\bp\]e^{-S_1-S_2}\cr=&\int\[Dz\]_{z_\d=0}
\[D\bp\]e^{-S_3[\bx_b,z,\bp]}=\int\[Dz\]_{z_\d=0}
\[D\bp\]e^{-S_0[\bp,1/z]+i\oint\bx_b\cdot d\bp}\cr
=&\int\[D\bp_b(s)\]e^{i\oint\bx_b\cdot d\bp_b}\int\[Dz\]_{z_\d=0}
\[D\bp\]_{\bp_\d=\bp_b}e^{-S_0[\bp,1/z]}~.}}

If we try to start in the opposite direction by considering an open
string in $AdS_5$ with boundary conditions such that it ends on the
curve $\CC_{\bx_b}$ at the Poincar\'e horizon $z=\infty$ then the
term
\eqn\boundaryt{\int_{\d\DD}d\sigma
z^2\d_\tau\bp\cdot\bp}
%
(which is now at $z=\infty$) is not zero (and diverges due to the
$z^2\to\infty$ factor). However, that term exactly cancels between
$\int_{\d\DD} d\sigma {\bf V}_\sigma\cdot\bp$ and $\int_{\d D}
d\sigma{\bf b}(\sigma)\int_0^\sigma ds\bV_\sigma(s)$ in \Buchertwo.

We have therefore seen, by using the identification of the
Wilson loop operator with open strings ending on the boundary of
AdS, that T-duality on the string worldsheet reproduces the Fourier
transform in loop space \MomentumWilson. As noted in \AldayHR, the
T-dual AdS space where the worldsheet fields are $\bp$ and $r$, has
a non trivial dilaton $\Phi\sim\log(r)$. Therefore, expectation
values of Wilson loops in momentum space are dual to open strings in
AdS ending on the Poincare horizon, where a dilaton is turned on in
the bulk, but no NS $B$ field is needed.\foot{This is a reference to
a suggestion of \PolyakovTJ.}

\newsec{Scattering amplitudes from worldline path integrals}

In this section we will derive a general relation between scattering
amplitudes of adjoint fields in $SU(N)$ gauge theories and Wilson
loops.\foot{Our derivation applies for any theory with an 't Hooft
limit.} The derivation will follow an example (gluon scattering
in pure YM theory), but holds for the scattering of any
adjoint fields in the planar limit.

Unless some couplings are scaled with $N$, fields that are not in
the adjoint representation
do not
contribute to planar scattering amplitude of adjoint fields.
Therefore, for our purpose, we can truncate the theory to the fields
in the adjoint only.

Let $\varphi^b_I(\bk)$ be the set of all fields in the adjoint
representation. These are the fields we wish to scatter. The index
$b$ is an $SU(N)$ adjoint color index, the index $I$ denotes the
field together with all of its labels, such as spin, flavor or other
global charge; $\bk$ is the momentum. Along with the general
discussion, we will follow the example where the only field in the
adjoint is the gauge field. In that case $I=\mu$ is a vector index.

We will want to include a source for $\varphi$, so that we may
scatter external $\varphi$ ``particles"; with this in mind, consider
the generating function
$$ Z[J] \equiv \vev{ e^{ \int J\cdot \varphi } }~,$$
where $J\cdot \varphi = J_b^I \varphi^b_I$. Note that if
$\varphi$ is fermionic, so is $J$.

To study on-shell scattering of $\varphi$ particles, we will use the
LSZ formula
\eqn\scatter{\AA_n\[(\bk_1,\varepsilon^{I_1}_{b_1}),\dots,(\bk_n,\varepsilon^{I_n}_{b_n})\]
= \prod_{i=1}^n \(\lim_{\bk_i^2\to
m^2}\varepsilon^{I_i}_{b_i}\(G_{ph}^{-1}(\bk_i)\)_{I_i}^{K_i}
 {\delta \over \delta J^{K_i}_{b_i}(\bk_i) }\) Z[J]|_{J=0}~,}
where $G_{ph}^{-1}$ is the fully-dressed propagator, $m$ is the
physical mass (so $G_{ph}$ has a pole at $\bk^2=m^2$), and we work
in the convention where all momenta are out-going.

For our YM example, in Feynman gauge
$$\lim_{\bk^2\to
m^2}\(G_{ph}^{-1}(\bk)\)_\mu^\nu=\lim_{\bk^2\to
0}h(\lambda)~\eta_\mu^\nu~ \bk^2~,$$
where $h(\lambda)$ is some function of the 't Hooft coupling
$\lambda$, and we used the fact that the Ward identity
protects the location of the pole of $G_{ph}$.

In the planar limit, a fixed color index is attached to any piece of
a planar diagram boundary between two
adjacent insertions.
As a result, for fixed number of external colored fields ($n$), we can split
the $N$ color indices as $N=n+M$, where $N,M\to\infty$ with $n$
fixed. The $SU(N)$ gauge group then naturally splits into
$SU(n)\times SU(M)$, where the amplitude transforms covariantly under
the global $SU(n)$ symmetry and is invariant the global $SU(M)$
symmetry (as well as the full $SU(N)$ local symmetry).

We split the fields accordingly as
$$\varphi=(a,w,A)~,$$
where $a$ are the fields that transform in the adjoint of $SU(n)$\foot{
We will reserve the non-boldface $a$ for the generic $SU(n)$ adjoint,
which are to be distinguished from the $\ba$s which will specifically
represent
the gauge fields in the YM example of section 6.},
$w$ are the fields that transform in the bi-fundamental of
$SU(n)\times SU(M)$ and $A$ are the fields transforming in the
adjoint of $SU(M)$. The source $J$ is then coupled only to $a$. Note
that this splitting of the fields is unambiguously defined only
asymptotically (where we set the states being scattered)
and is gauge-dependent
in the bulk. Therefore, it requires partial gauge fixing
(as will be
done for our example in section 6), and some of the $w$ fields are
the corresponding ghosts.
This gauge-fixing leaves an $SU(n)\times SU(M)\subset SU(N)$
subgroup unfixed. 

At large $N$, the $a$ field contributes only at tree level. Any $w$
field goes only on the boundaries of planar diagrams and therefore
contributes only at one loop to one-particle irreducible (1PI)
diagrams. We will incorporate these simplifications by first
computing all planar 1PI diagrams bounded by $w$ and then connecting
these into trees with $a$ propagators (see fig.\ 1).

\fig{A generic (reducible) planar diagram contributing to the
scattering amplitude of five $a$ fields.}{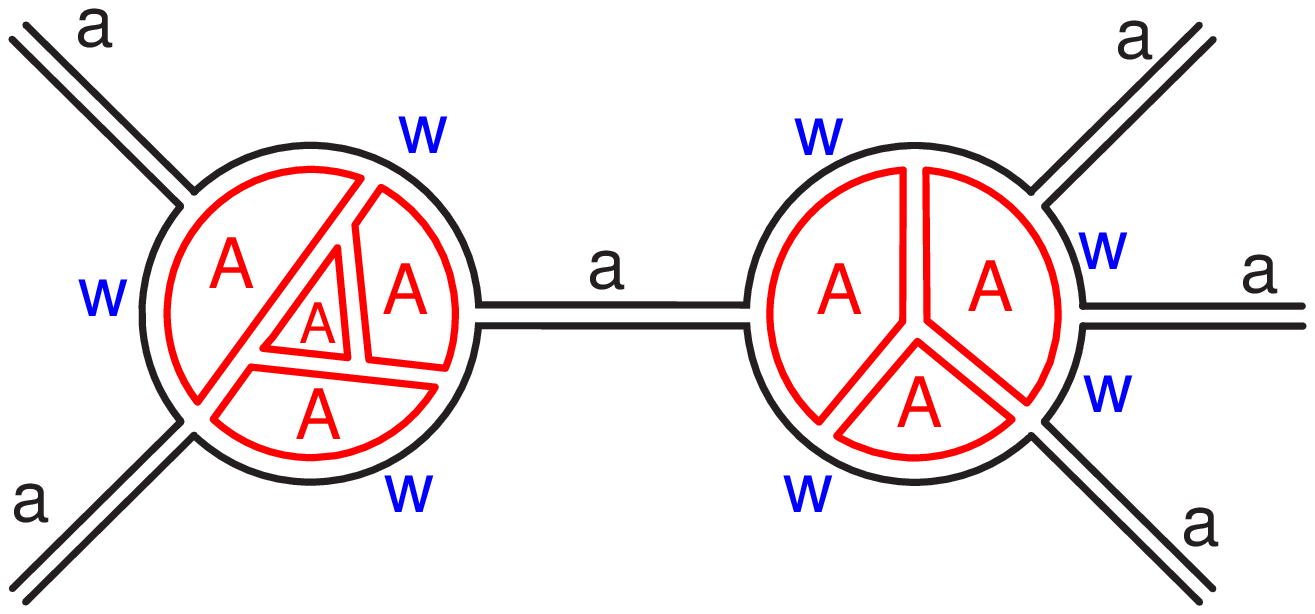}{2.4in}

To do so, we first add auxiliary fields transforming in the adjoint
of $SU(n)$ to express all interactions involving four $w$'s (and in
general higher) as interactions of only two $w$'s with the auxiliary
fields. For instance, in our pure YM example there is a non-abelian
interaction involving four $w$-bosons. As will be explained in
section 6, we first express it as (see fig.\ 2)
\eqn\fourvertex{\eqalign{\tr_{n\times
n}\(w_{[\mu},w^\dagger_{\nu]}w^{[\mu}{w^{\nu]}}^\dagger\)
\quad\to&\quad 2\tr_{n\times
n}\(w_{[\mu}w^\dagger_{\nu]}d^{\mu\nu}\)-\tr_{n\times
n}\(d^{\mu\nu}d_{\mu\nu}\)\cr \tr_{M\times
M}\(w^\dagger_{[\mu},w_{\nu]}{w^{[\mu}}^\dagger w^{\nu]}\)
\quad\to&\quad 2\tr_{M\times
M}\(w^\dagger_{[\mu}w_{\nu]}e^{\mu\nu}\)-\tr_{M\times
M}\(e^{\mu\nu}e_{\mu\nu}\)~,}}
\fig{An auxiliary field is added to open up a four-$w$
vertex.}{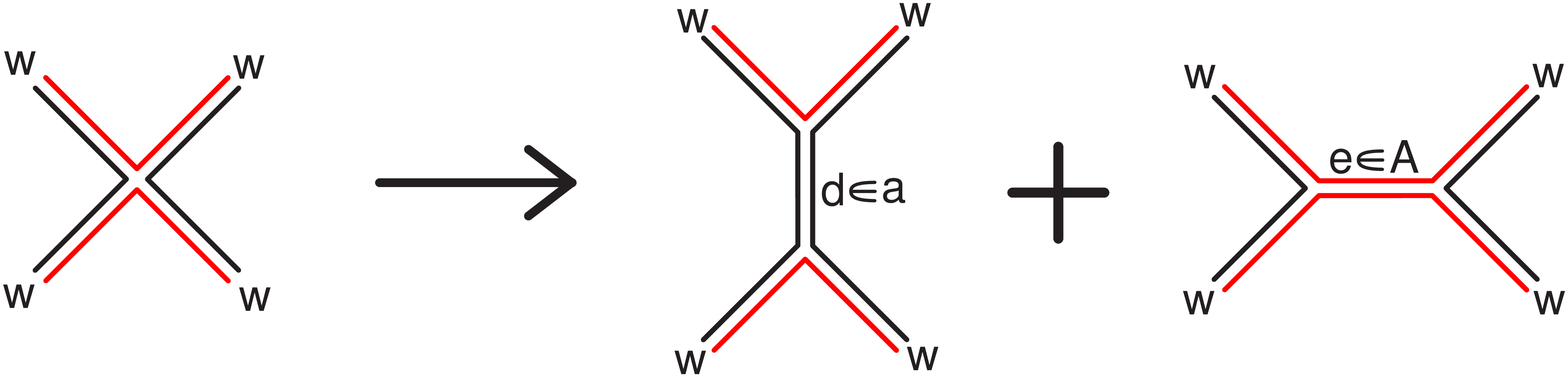}{3.6in}
\noindent where $w_\mu=w_\mu^b T^b_{n\times M}$ and $d\in a,~e\in A$
are anti-symmetric auxiliary tensor fields. The arrows indicate
classical equivalence, which is exact at large $N$ (for fixed number
of external particles).\foot{Instead of introducing these two types
of auxiliary fields, we could express the two vertices in
\fourvertex\ with a single auxiliary field in the adjoint of $SU(n)$
only ($d^{\mu\nu}\in a$). Such an auxiliary tensor field would not
be anti-symmetric. Both types of auxiliary fields have advantages
and disadvantages that will be discussed in section 6. Here, for
simplicity, we have chosen the first option.} The resulting action is
now quadratic in $w$. Integrating it out results in a one-loop
determinant, and the generating function is now
\eqn\genoneloop{Z[J] = \vev{\det\({\delta^2 S\over\delta^2
w}\)_{w=0}e^{ \int J\cdot a }}_{a,A}~.}
As will be derived in section 6, for our pure YM example in Feynman
gauge
\eqn\YMdet{\det\({\delta^2 S\over\delta^2
w}\)_{w=0}=\det\[-D^2\]\det\[-D^2\eta+i2F-d -e \]^{-\half}~.}
If the scattered partons are massless, then there are IR divergences
which should be regularized. In section 6 we add a Higgs field
which gives the $w$-boson a mass,
in order to regulate these IR divergences
in the gluon amplitude. Our formal
derivation here will rely only on large $N$, and for
massless partons different IR regulators may be used.

Next, we express the one loop determinant using the worldline
formalism. That should be possible at least for any gauge theory that
can be obtained as the low energy limit of some open string theory. In
such a string theory description, the adjoint fields are represented by
open strings stretched between D-branes. The field theory one loop
determinant is then obtained from the string theory as the low
energy limit of the annulus diagram in which the string length goes
to zero. In that limit, the annulus becomes a circle and the string
worldsheet
theory becomes a worldline theory on the circle. The one loop
determinant in \genoneloop\ is coupled to arbitrary background
fields. When coupling an open string to a background field, the
worldsheet conformal symmetry restricts the background to be
on-shell. In the worldline limit however, there is no two
dimensional conformal symmetry and one can consistently take the
background off shell.\foot{For the Brink-Schwarz and pure spinor
worldline descriptions of 10d $\NN=1$ or 4d $\NN=4$ SYM theories, the
background is restricted to be on-shell. This is, however, a
technical complication resulting from the absence of an off-shell
superspace description of these gauge theories. It may be resolved
by a harmonic superspace worldline description of these theories
\BuchbinderUB.}

As we will review in the next section and the appendices,
the worldline representation of the one loop determinant is of the
following form
$$\det\({\delta^2 S\over\delta^2
w}\)_{w=0}=\exp\(\int{dT\over T}\NN\int[D\bx]_1[D\xi]\tr
Pe^{iS_{wl}[\bx,\xi;a,A]}\)~,$$
where $\bx(s)$ is the worldline path, $\xi$ stands for all other
worldline fields, $S_{wl}$ is the worldline action and $P$ stands
for path ordering. Since the derivatives in the one loop determinant
are covariant derivatives, the worldline action will contain a
linear coupling to the gauge field
$$\oint (\bA-\ba)\cdot d\bx~\in~ S_{wl}~,$$
where the coupling of $\bx$ to $\ba$ and $\bA$ have opposite signs,
since $\bw$ is in the $({\bf n}, {\bf \bar M})$ representation. As
such we will call the path ordered exponent of the worldline action
a {\it generalized Wilson loop} and the worldline path integral a
{\it Wilson sum}.

The worldline representation of the YM example
\YMdet\ is given in section 6 and the appendices, following
\StrasslerZR. In that description we have a sum over two worldline
path integrals, one representing the vector determinant and the
other representing the ghost determinant (which therefore has a
minus sign in front). That worldline representation may be
considered a gauged-fixed version of a single worldline path
integral with some local (super) symmetries, where the ghost
determinant piece comes from some worldline ghosts. The reason it
broke up into a sum of two worldline path integrals is that these
are the one-particle worldline theories standing in the exponent and
we don't have background ghost, as in tractable string worldsheet
theories. That is because \YMdet\ will be obtained without fixing a
gauge for the $SU(n)\times SU(M)$ part of the gauge group. As we
will scatter the $SU(n)$ fields, we will fix a gauge for that part
as well and add the corresponding ghost as part of the $a$ fields.
However, since we do not scatter these $SU(n)$ ghosts (which, by
large-$N$, contribute only at tree level), ghost number conservation
implies they will not contribute in the planar limit and will be
harmlessly set to zero.

Next, we split the gauge theory path integral into an explicit path
integral over $a$ (from which only the tree level contributes at
large $N$) and a path integral over $A$ written in terms of
expectation values $\<...\>_A$. The resulting generating functional
of correlation functions of $a$s is
\eqn\genexp{Z[J]=\int[Da]e^{ \int \(S[a]  + J \cdot a \)}
\left\langle\exp\(\int{dT\over T}\NN\int[D\bx][D\xi]\tr P
e^{-S_{wl}[\bx,\xi;a,A]}\) \right\rangle
_A~,}
where $S[a]$ is the $SU(n)$ gauge theory action (which includes the
auxiliary field couplings in \fourvertex).  In \genexp\
and in the remainder of this section, by $[D\bx]$ we mean
the measure $[D\bx]_1$,
defined in \dxone.

For our example, in Feynman gauge, the piece in \genexp\ outside of
the $\<...\>_A$ expectation value is
$$\int[Da]~e^{ \int \(S[a]  + J \cdot a\)}
\dots =\int[D\ba][Dc][Dd]e^{ \int \tr\(\half \ba\cdot \square \ba
+\bar c\square c+ \bJ\cdot \ba-d^2/4\)  } \dots~~~.$$

Up to this point, we have not used the large $N$ limit and our
expression for the generating function \genexp\ is exact for any
value of $N$. We now use large $N$ factorization of Wilson loop
expectation values to lift the $A$-expectation value into the
exponent
\eqn\genexp{Z[J]=\int[Da]\exp\( \int \(S[a] + J \cdot a
\)+\int{dT\over T}\NN\int[D\bx][D\xi]\<\tr P
e^{-S_{wl}[\bx,\xi;a,A]}\>_A\)~.}
\fig{A class of non-planar diagrams that are down by ${1\over M}$
and are removed in \genexp.}{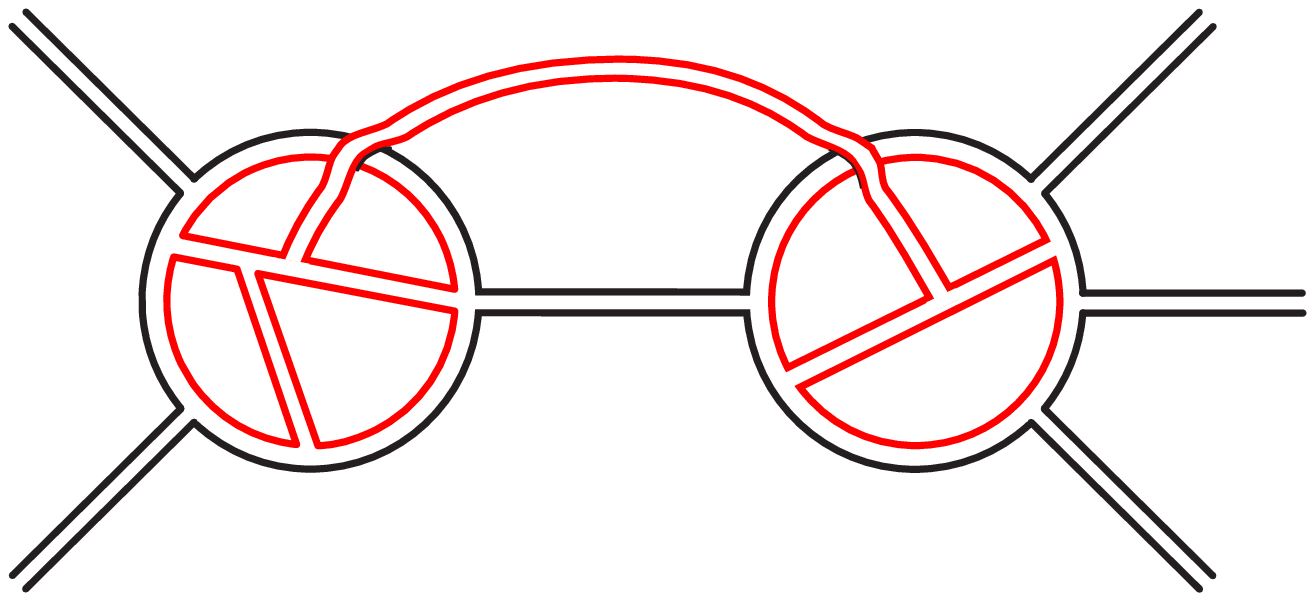}{2in}
\fig{A class of planar diagrams that are down by ${n\over M}$ and
are removed by keeping only the tree level contributions in
$a$.}{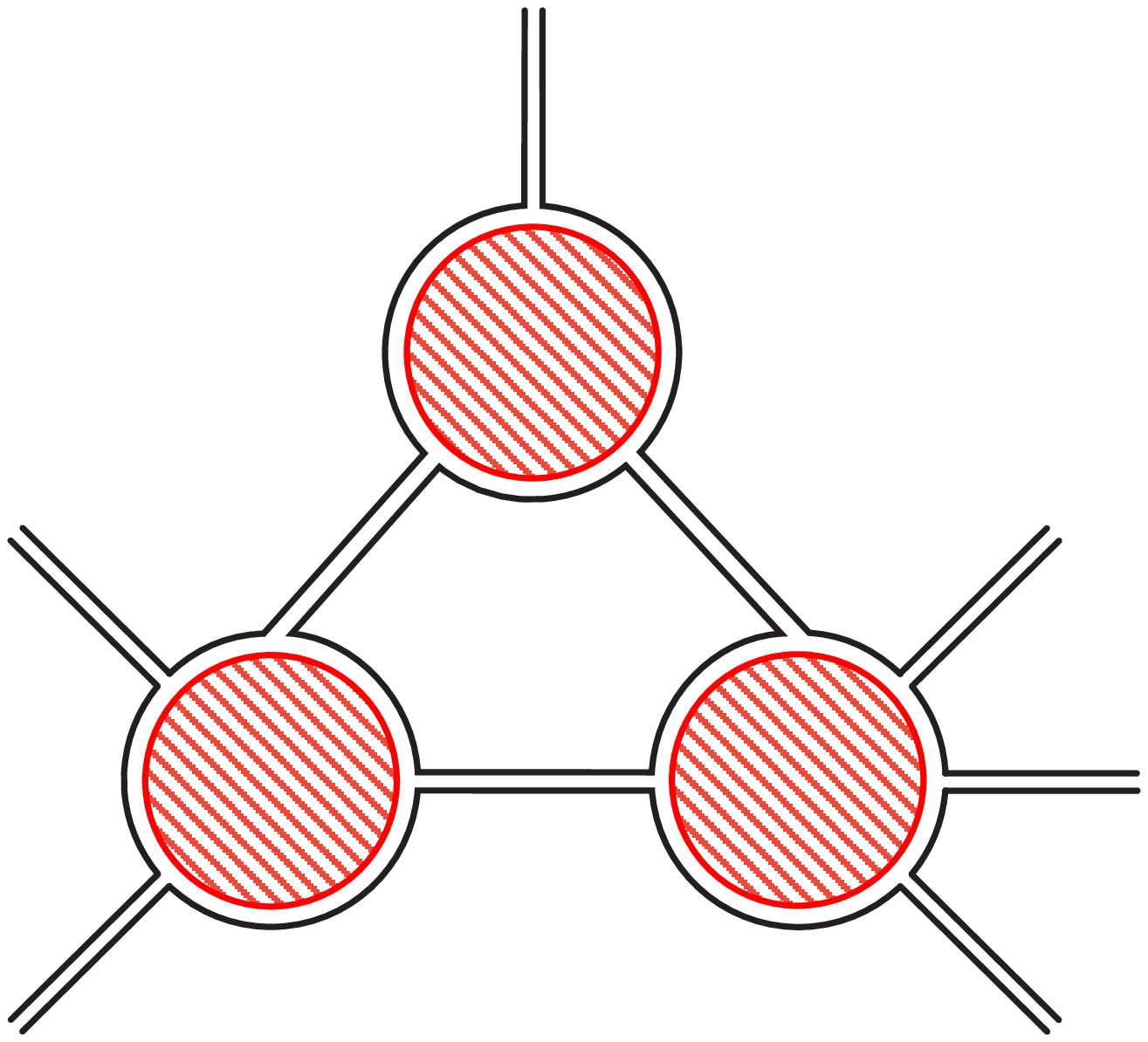}{2in}
In writing equation \genexp\ we have removed a class of (non-planar) diagrams that
are down by ${1\over M}$ and are shown in figure 3. There are also
diagrams contributing to \genexp\ that are down by powers of
${n\over M}$ (planar and non-planar) (see fig.\ 4). In a diagrammatic
expansion of the path integral over $a$, the generalized Wilson
loop expectation values
$$ \<\tr P e^{-S_{wl}[\bx,\xi;a,A]}\>_A$$
play the role of vertices of any order (which are not local in
space). In that diagrammatic expansion, we need to keep only the tree level
contributions. This amounts to keeping only the leading contribution
in ${n\over M}$ since the color index in an $a$ loop runs only over
$n$ indices (see fig.\ 4). In other words, the planar 1PI {\it quantum
effective action} of the $a$ fields is\foot{Note that the last
(worldline) term can also contribute to the mass of $a$,
in the case that it is not protected by a Ward identity.}\foot{The
structure of the non-planar corrections is very interesting but we
postpone their discussion.}\foot{A formula of this form,
incorporating only the contributions of scalars and fermions
in $w$ (but not $w$-bosons), appears in \S6 of \BrandtGZ.}
\eqn\qea{\Gamma[a]= S[a] +\int{dT\over T}\NN\int[D\bx][D\xi]\<\tr P
e^{-S_{wl}[\bx,\xi;a,A]}\>_A}
and the sum over connected diagrams with sources is
$$W[J]= S[a_J]+\int J\cdot a_J +\int{dT\over
T}\NN\int[D\bx][D\xi]\<\tr P e^{-S_{wl}[\bx,\xi;a_J,A]}\>_A~,$$
where
$$-J={\delta\over\delta a}\(S[a]+\int{dT\over
T}\NN\int[D\bx][D\xi]\<\tr P
e^{-S_{wl}[\bx,\xi;a,A]}\>_A\)_{a=a_J}~.$$
In particular, for two separated points $\by\ne \bz$, the
fully-dressed inverse propagator is given by
\eqn\dressed{G_{ph}^{-1}(\by,\bz)_{b_1b_2}={\delta^2\over\delta
a(\by)\delta a(\bz)}\[S[a]+\int{dT\over T}\NN\int[D\bx][D\xi]\<\tr\(
P e^{-S_{wl}[\bx,\xi;a,A]}\)\>_A\]_{a=0}\delta_{b_1b_2}~.}
When an external leg is dressed by $G_{ph}$ \dressed, we get a pole
which is then canceled by $G^{-1}_{ph}$ in \scatter. The 1PI
$m$-vertices $V_m$ with $m\ge 5$ are given by
%
\eqn\mvertex{\eqalign{&V_m\[(\bk_1,\varepsilon_{b_1}^{I_1}),
\dots,(\bk_m,\varepsilon_{b_m}^{I_m})\]\cr=&\prod_{i=1}^m\int
d\bx_ie^{-i\bk_i\cdot
\bx_i}\varepsilon_{b_i}^{I_i}{\delta\over\delta
a^{b_i}_{I_i}(\bx_i)}\int{dT\over
T}\NN\int[D\bx][D\xi]\<\tr_{N\times N}\( P
e^{-S_{wl}[\bx,\xi;a,A]}\)_{a=0}\>_A\cr =&\sum_\pi \tr_{n\times
n}\(T^{b_{\pi(1)}}\dots T^{b_{\pi(m)}}\)\int{dT\over
T}\NN\int[D\bx][D\xi]\prod_{i=1}^m\int_{s_{\pi(i-1)}}^{s_{\pi(i+1)}}
ds_{\pi(i)}\varepsilon_{b_i}^{I_i}V(s_1)_{I_i}e^{-i\bk_i\cdot
\bx(s_i)}\cr &\times\<\tr_{M\times M}\(P
e^{-S_{wl}[\bx,\xi;0,A]}\)\>_A~,}}
where the sum is over all permutations, $s_{\pi(0)}=0$,
$s_{\pi(m+1)}=T$ and
$$\varepsilon_{b}^{I}T^{b}V_{I}(s)=-{\delta\over\delta
a^b_I(\bx(s))}S_{wl}[\bx,\xi;a,A]_{a=0}$$
is the worldline vertex operator.\foot{Note that in case where the
loop has self-crossings, the variation with respect to the source at
the crossing point will give a sum of the two possible orderings.}
For $n\le 4$, in addition to \mvertex\ one also have to include the
$a$-tree level vertex. In our example when $a=a^\mu_b$ is the gauge
field, we have
$$\varepsilon_b^\mu V_\mu(s)T^b=\varepsilon_b^\mu\dot
x_\mu(s)T^b$$
from the scalar worldline and
$$\varepsilon_b^\mu V_\mu(s)T^b=\varepsilon_b^\mu\(\dot
x_\mu(s)+ik_i^\nu[\psi_\nu,\bar\psi^\mu]\)T^b$$
from the vector worldline, where $\psi^\mu$ is a complex worldline
fermion field (see section 6).

In \mvertex\ we assumed that $S_{wl}$ is linear in $a$. However, in
our example in section 6, $S_{wl}$ will have quadratic coupling to
$a$. In our example, the vector worldline contains the non-abelian
coupling
\eqn\quadratic{S_{wl}=\int ds\psi^\mu\bar\psi^\nu
F_{\mu\nu}+\dots=\int ds\psi^\mu\bar\psi^\nu[a_\mu,a_\nu]+\dots~.}
This term in $S_{wl}$ is necessary for the gauge invariance of the
operator
$$\int[D\bx][D\xi]\tr Pe^{-S_{wl}[\bx,\xi;a,A]}~.$$
It leads to what is called
the {\it two gluon vertex}
in the formulation of the theory
in terms of worldline path integrals
\StrasslerZR.
In general it reads
$$\varepsilon_b^I\varepsilon_c^J T^b T^c V_{IJ}(s)=
{\delta\over\delta a^b_I(\bx(s))}{\delta\over\delta
a^c_J(\bx(s))}S_{wl}[\bx,\xi;a,A]_{a=0}~.$$
The corresponding vertex is indicated in figure 5b. These contribute
only at the boundary of the integration over the vertices
insertion points (see figure 5a).\foot{In a superspace
representation of the worldline path integral, the path ordered
exponent is replaced by a super-path-ordered exponent (see
\AndreevBZ\ for an $\NN=1$ example). In that representation of the
worldline 1PI vertex there are no two-gluon vertices and the
corresponding boundary contact terms result from the super-path-ordering.}
\fig{{\bf a.} A boundary point in loop space where a sub-loop between
two adjacent insertions collapses to a point. {\bf b.} A $w$-loop
interacting with two $a$ fields at a four-vertex.}{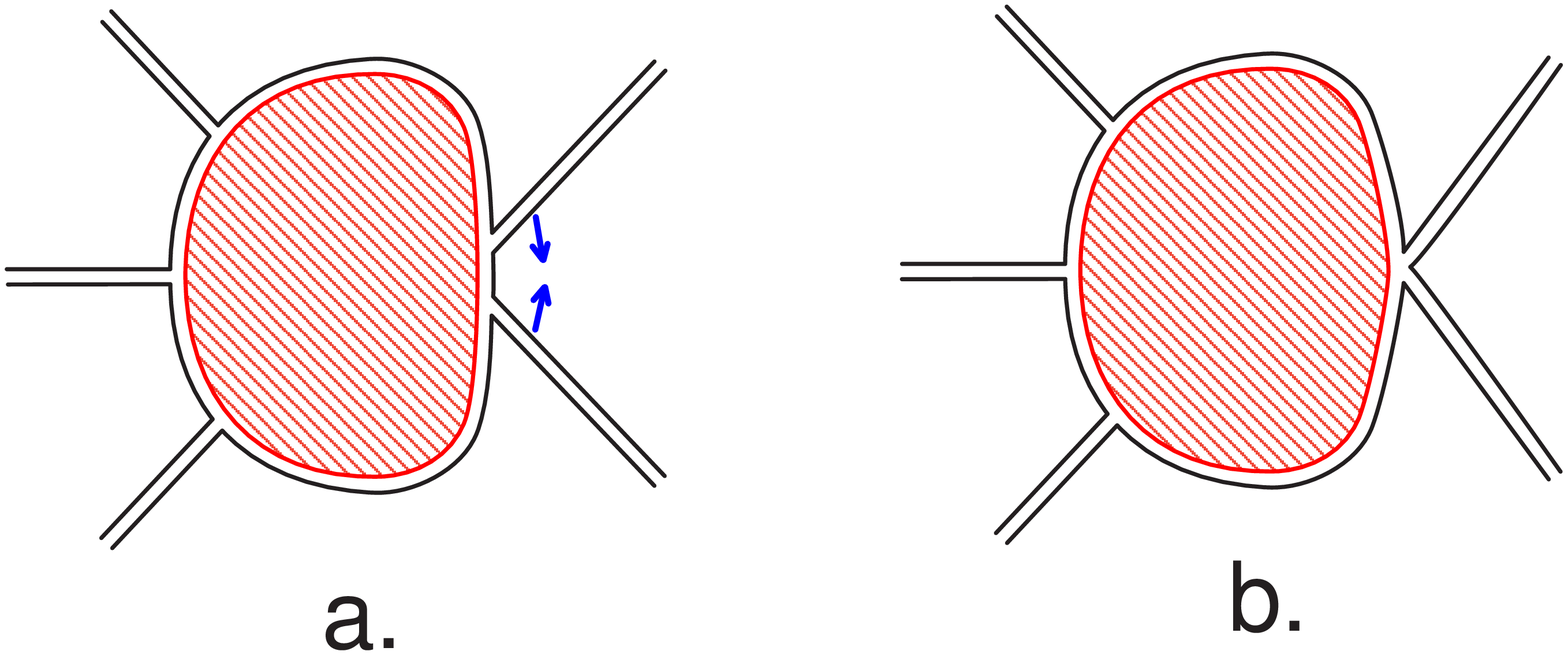}{3.2in}
In our example, by partial integration, the coupling to the field
strength $F_{\mu\nu}$ can be replaced by
\refs{\Migdal,\BrezinSK,
\AvramisXF}
%
\eqn\spinfactor{\int ds\psi^\mu\bar\psi^\nu
F_{\mu\nu}~\to~\int ds\psi^\mu\bar\psi^\nu \dot x_{[\mu}\ddot
x_{\nu]}~.}
Using this `radiation-reaction term,'
the two-gluon vertex can be avoided.

Similarly, since the auxiliary fields couple only linearly to the
worldline action, are never external, and have no kinetic terms,
integrating them out leads to an interaction between position space
Wilson loops only when these touch at a point. As can be seen in
figure 6, these are in one-to-one correspondence with points in the
path integral over closed loops where the loop is self-crossing.
Therefore, these contributions can be represented by a correction to
self-crossing points of generalized Wilson loops. We can represent
these corrections by replacing the generalized Wilson loop
expectation values with
\fig{{\bf a.} Two Wilson loops representing two $w$-loops
interacting at a four-vertex. {\bf b.} A Wilson loop with a crossing
point representing a single $w$-loop.}{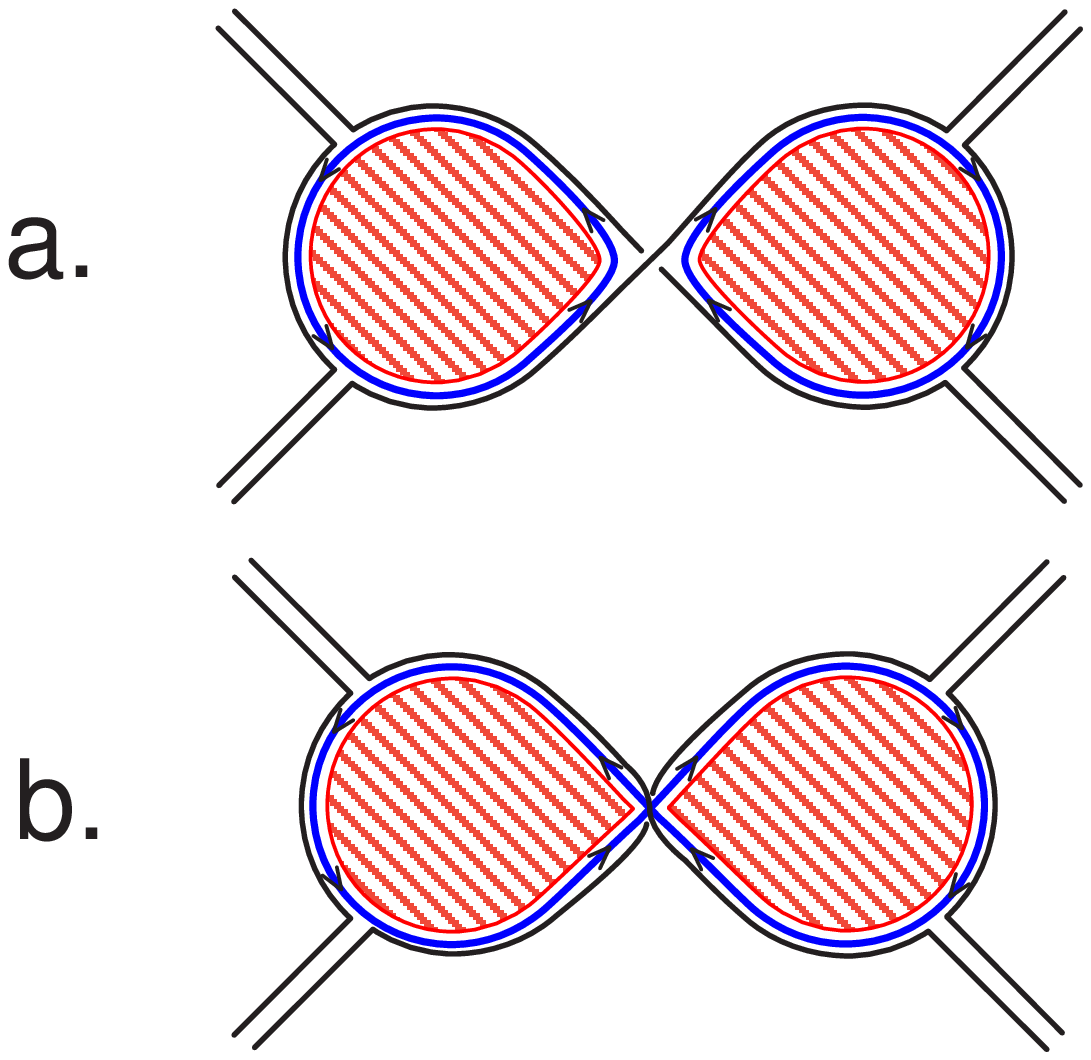}{2.2in}
$$\<\tr P e^{-S_{wl}[\bx,\xi;A]}\>_A^{s.c.}~,$$
where $s.c.$ stands for {\it self-crossing correction}. Note that
these
corrections are localized on the loop and
contribute at $\lambda^1$ only. Therefore, they can be computed
perturbatively. We leave such computation to future work.


The momentum dependence of any 1PI $m$-vertex $V_m$ is
$$e^{-i\sum_{i=1}^m\bk_i\cdot \bx(s_i)}=e^{i\oint \bp\cdot d\bx}~,$$
where
\eqn\gpolygon{\bp(s)=\sum_{i=1}^m \bk_i\theta(s-s_i)~.}
Therefore, each planar 1PI $m$-vertex is a {\it generalized momentum
Wilson loop}.\foot{Note that in the expression for the 1PI vertex,
it is only after we integrate over the vertices insertion points
that we find a meaningful, parametrization-invariant
functional of the gluon momenta and
polarizations.} In momentum loop space, the large $N$ factorization
of the partial amplitude into 1PI planar gluonic blobs connected by
(dressed) propagators has a geometric manifestation as a sum over
all the ways of subdividing the polygon made of the external momentums
into sub-polygons connected by the
$a$
propagators (see fig.\ 7).\foot{A
very similar picture was used in \DrummondAU\ when realizing the
dual conformal symmetry in $\NN=4$ perturbation theory.}
\fig{A one-particle-reducible contribution to the scattering amplitude
is a subdivision of the polygonal momentum wilson loop.
}{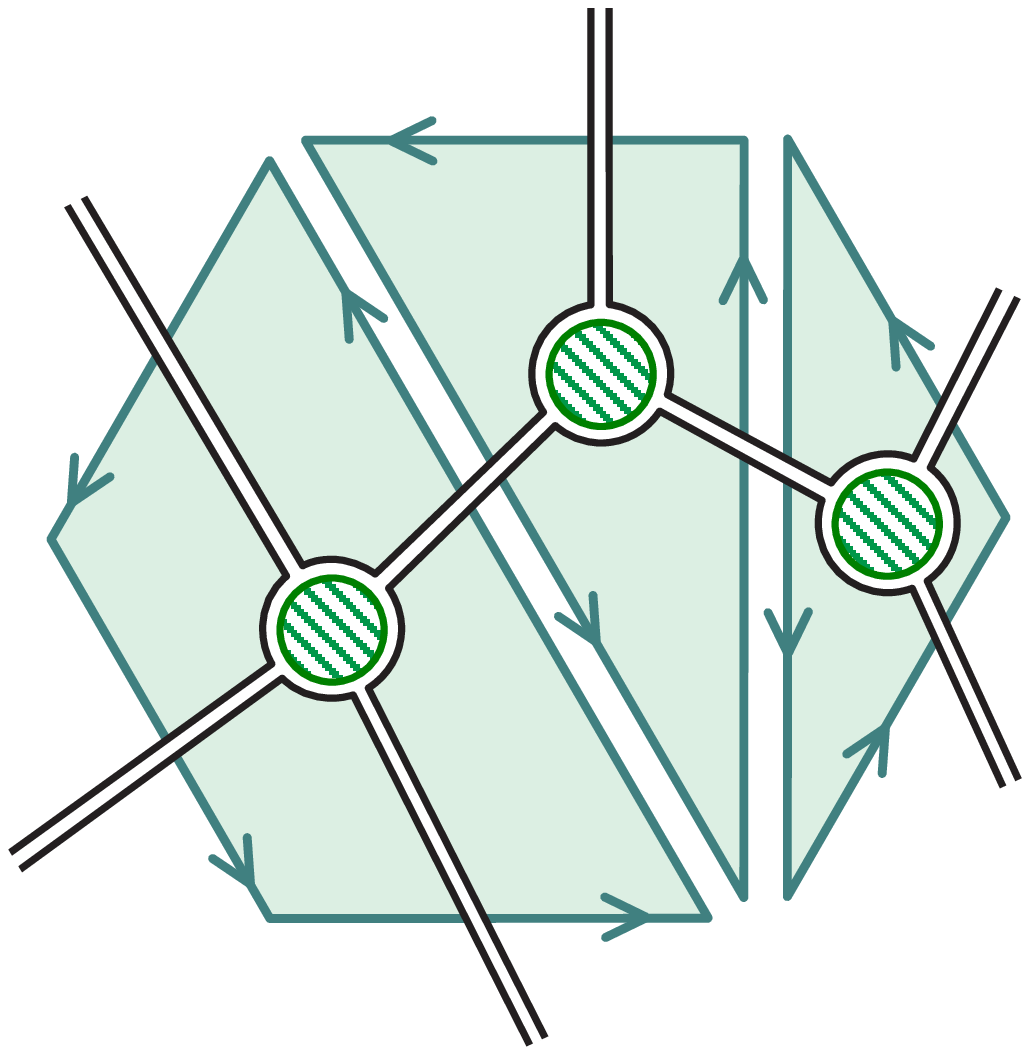}{2in}

The expressions for the 1PI planar vertices and the propagator in terms
of
generalized Wilson loop expectation values represent a non-perturbative
definition of these quantities. For the scattering amplitude of any
finite number of partons ($n$), there is a finite number of tree
level diagrams connecting the external legs. For 't Hooft coupling
of order one, all of them contribute and should be summed over. The
large 't Hooft coupling limit will be discussed in the next section.
In that limit we will argue that for generic external momenta, the
single 1PI $n$-vertex will dominate the sum. In momentum space it is
given by a single polygon made of the ordered external momenta.

\newsec{One-particle reducible contributions}

In perturbation theory, there are non-1PI (one-particle reducible)
contributions that are not suppressed by a power of $1/N$. Here, we
wish to understand what is known about the contribution of this
class of diagrams outside of perturbation theory.

In string theory, the moduli space of the disk with $n$ insertion
has boundaries.
The components of this boundary are in one-to-one correspondence
with the non-1PI diagrams. A separation between 1PI and non-1PI
contributions is not gauge invariant. Similarly, in the string
theory, restricting the integral over gluon vertex operators
insertion points to some specific point away from the saddle point
is not gauge invariant and depends on the specific worldsheet
formalism being used. In any given worldsheet formalism, we suggest
to identify the contribution to the string theory amplitude from the
boundary of moduli space with the non-1PI contributions to the gauge
theory amplitude in some specific gauge corresponding to the given
worldsheet formalism. We will then use that identification to argue
that at strong coupling, in any gauge, the 1PI contributions
dominate the amplitude.

\fig{
A one-particle reducible
(\ie\ non-1PI)
diagram (b) and the corresponding
boundary point in the disk moduli space (a).}{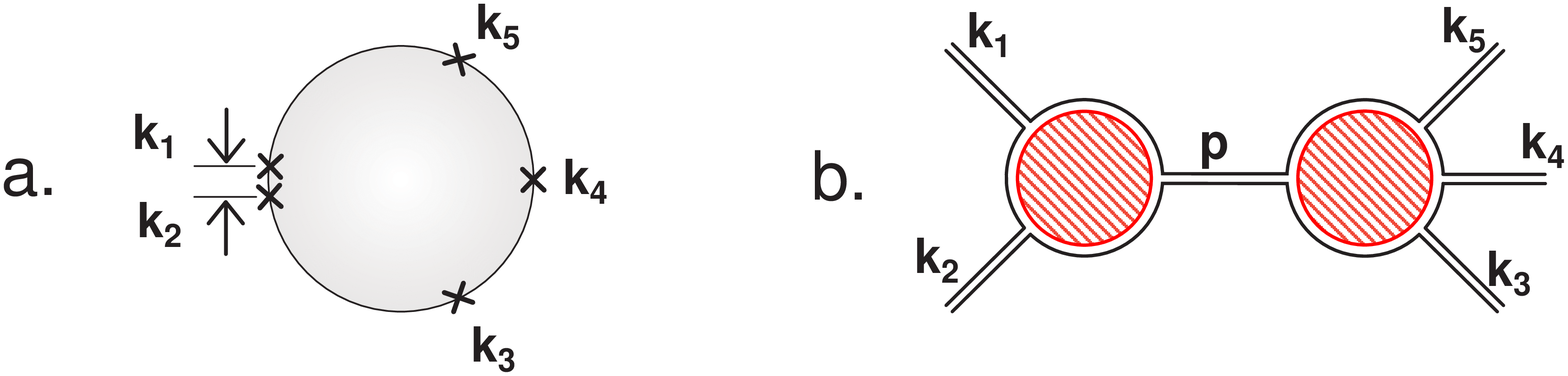}{4in}

Alday and Maldacena \AldayHR\ argue that the gluon scattering
amplitude at large $\lambda$ is dominated by a saddle point which
for generic external momenta sits away from the boundary of the
moduli space. Therefore, assuming the identification above, the
leading contribution to the amplitude cannot come from one-particle
reducible diagrams at strong coupling.

In addition, if we restrict the integral to one of the boundary
components (see fig.\ 8), and look for the saddle point using the AM
holographic prescription we find a contribution which is indeed much
smaller than that of the leading saddle point. This had to be true
given the previous statement, since we are here extremizing over a
subset of the original set; unless the extremum of the bigger
integral lies in the subset, the inequality will be strict.

We would like to use this fact to argue that the 1PI contributions
dominate at large 't Hooft coupling. Specifically, the non-1PI
contributions to our formula are expressed as a product of
amplitudes each with one off-shell external leg, attached by a
dressed gluon propagator with momentum $ p \equiv k_1 + ... + k_r$.
At large $\lambda$, we can try to compute each factor by a worldsheet
saddle-point calculation.
For generic $\{k_1, \dots, k_r\}$,
the area of the resulting worldsheet
gets a divergent contribution from the region where it attaches to this
non-null edge at the boundary.
Heuristically, this is the tension of the flux tube carried by the off-shell gluon
through the strongly-interacting theory.\foot{More speculatively, this suggests a possible holographic description of
off-shell amplitudes.  However,
we warn the reader that have not given a gauge-invariant
definition of the off-shell scattering amplitude,
nor have we given a reparametrization-invariant definition
of the polygon momentum Wilson loop.}

For special values of the external momenta,
the saddle point of the big integral {\it does} lie
on the boundary of moduli space.
In this case, the intermediate gluon which
connects the two sub-diagrams is on shell.
In this case, the intermediate line is null,
just like all of the external lines, and the
contribution to the worldsheet area is small
away from the cusps.

Note that there are two competing limits here: for any $ \bp^2 \neq
0$, the $e^{ - \sqrt \lambda }$ pushes the saddle point away from
the boundary of moduli space, while for $ \bp^2 \sim 0$, the pole in
the propagator gives a large contribution.\foot{Note that at
$\bp^2=0$, the factorized polygon has more cusps than the
un-factorized one and is therefore more suppressed by its Sudakov
form factors. However, the IR cutoff is never removed and therefore
at $\bp^2=0$ the pole in the intermediate propagator dominates.}
Therefore, if we first take $\lambda \to \infty$ before taking the
collinear limit $\bp^2 = 0 $, we will miss this factorization
contribution. \foot{The mysterious factorization behavior was
noticed by Chung-I Tan. We thank him for raising this question to
us.}

For finite values of the 't Hooft coupling, we have derived a
formula for the scattering amplitude as a sum over all subdivisions
of the polygonal momentum loop. There is by now some evidence that
the scattering amplitude is equal to the polygonal Wilson loop in
position space, both at strong coupling \AldayHR\ and at weak coupling
\refs{\DrummondAU, \BrandhuberYX, \DrummondAQ, \BernAP}. This conjecture was suggested by the similarity between
position and momentum loops following from T-duality in AdS \AldayHR.
However, we now see that at finite coupling, the single momentum
loop cannot equal the scattering amplitude. In particular, the
combination of these two formulae for the scattering amplitude
(the sum
over momentum polygon subdivisions and the position space polygon) would
imply that the polygon in position space is equal to the sum over
subdivisions of the polygon in momentum space!

We leave the further exploration of this suggestion to future work.

\newsec{Example: Yang-Mills with adjoint Higgs}

In this section we work out
in an example explicit representations for
the worldline integrals written more generally and
abstractly in section 4.
We proceed in two steps, first (in section 6.1)
identifying the fields mediating the planar scattering (
collectively called $w$ in the notation
of section 4) in an unambiguous way,
and writing their
contribution in terms of products of determinants,
and then (in section 6.2) representing these determinants with worldline path
integrals.

\subsec{Integrating out the bifundamentals}

In this section we will realize the general planar structure
obtained in section 4 for a specific example. The
simplest example is the scattering of gluons in pure $SU(N)$ YM
theory and we will start by describing that. However, since gluon
are massless, their amplitudes are suppressed by IR-divergent
Sudakov form factors. To regularize these IR divergences we will add
a real adjoint Higgs field. The Higgsed theory can most simply be
obtained by a KK reduction from 4+1 dimensions.\foot{
In this non-supersymmetric theory, in the absence
of a potential for the Higgs, radiative corrections will push the
Higgs vev back to zero. We ignore these below.} The Higgs will then
be used to give a mass to some of the gluons and thereby to regulate
these IR divergences. Note however that we will not scatter the
corresponding $w$-bosons. These will run only on the boundary of
planar diagrams and therefore will regulate the IR divergences
coming from planar diagrams. The same kind of regulator was also
used by Alday and Maldacena in their holographic description of the
scattering amplitude \AldayHR. In that holographic description, the
massive $w$-boson arises from separating the finite stack of $n$ D3
branes to which the asymptotic gluons are attached (see fig.\ 9).
\fig{A generic planar diagram that contributes to the gluon
scattering amplitude; the IR divergences are regularized by
separating the corresponding D3-branes.}{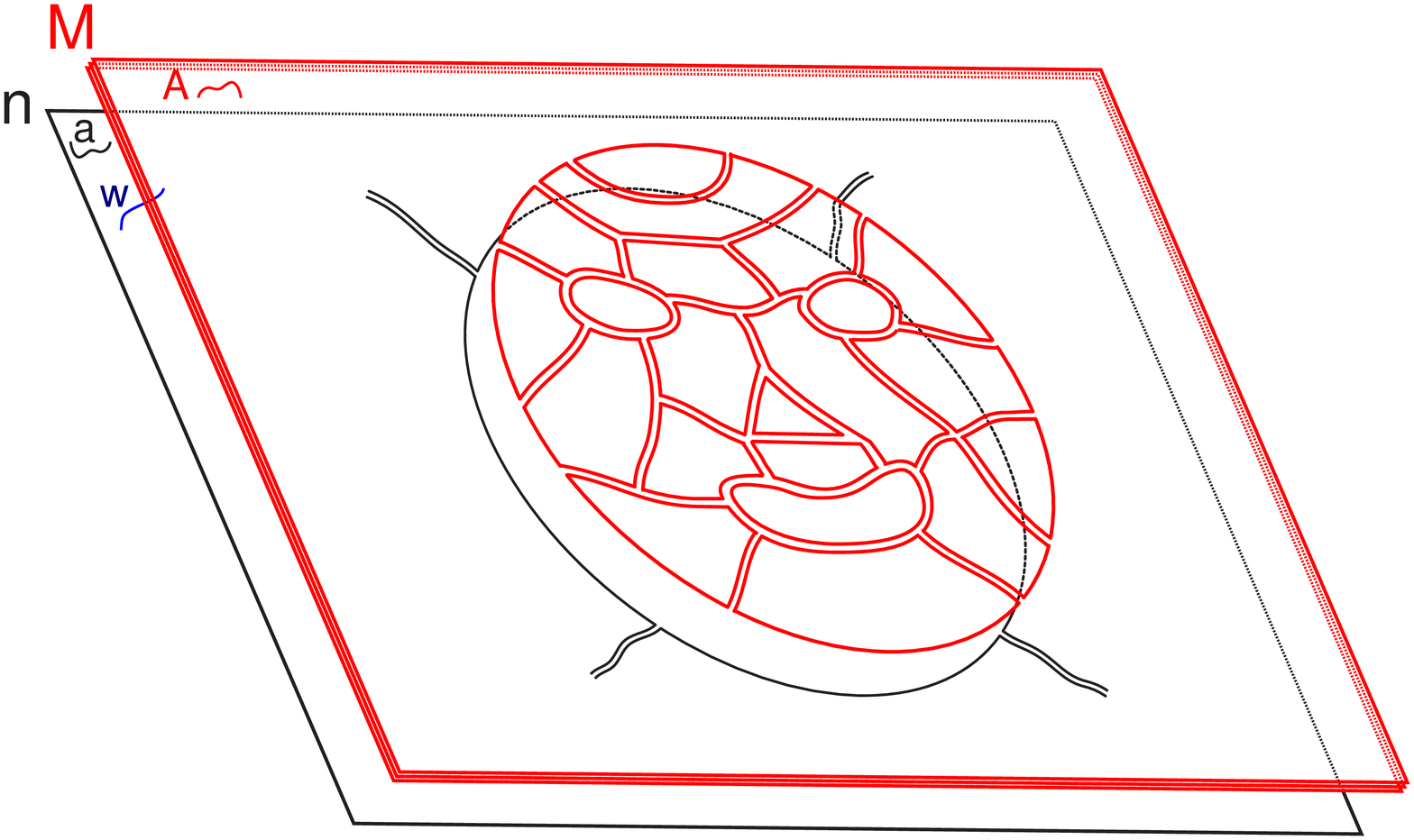}{3.4in}
This regularization has a natural description from the worldline point of view.
Most studies of
perturbation theory, however, use dimensional
regularization. A realization of dimensional
regularization in the
worldline description of the gauge theory
is suggested in \Schubert.

We start from $SU(N)$ gauge group where $n\ll N$ is the number of
color indexes carried by the asymptotic gluons. We then Higgs the
theory to $SU(n)\times SU(M)$, where $M=N-n$.

We will be interested in the large $N$ 't Hooft limit, keeping $n$
finite. As explained in previous 4, in that limit, loops on
which the color index runs over the $SU(n)$ part of the gauge group
are down by $n/N$. Therefore in the planar limit, the $\ba$
fields contribute only at tree level, whereas the $\bw$ fields
contribute only at one loop to 1PI diagrams.

We start with the pure $SU(N)$ YM theory of a gauge field
$\widetilde\bA$. To fix a gauge, we add to the Lagrangian
the gauge-fixing and corresponding ghost terms
\eqn\gfixterms{\LL_{gf}+\LL_{gh}=-\half \tr\(G^2\)-\bar c^a{\delta
G^a\over\delta\theta^b}c^b~,}
where $G$ is a gauge-fixing function and $c$ are the Faddeev-Popov
ghosts. To choose a gauge-fixing function that only partly fixes the
gauge, we introduce the matrix
\eqn\Higgs{v=\(\matrix{\One_{n\times n}&\ \cr\ & 0_{M\times M}}\)~.}
We then choose a Feynman-like gauge
$$G=[v,[v,\widetilde\bD\cdot\bw]]$$
where
$$\bw=[v,[v,\tilde\bA]]$$
and by taking two commutators with $v$ we have projected on the
piece that does not commute with $v$.

In that gauge, the gauge field splits as
\eqn\decompose{\widetilde\bA=\(\matrix{\ba_{n\times n}&\bw_{n\times
M}\cr \bw^\dagger_{M\times n}&\bA_{M\times M}}\)~;}
it will be clear from the context when fields are full
$N\times N$ color matrices or their corresponding sub blocks
\decompose. In this notation,
\eqn\gaugefix{G=\bD\cdot \bw=\d_\mu w^\mu-i\[A_\mu+a_\mu,w^\mu\]~.}

Let $\{\theta_a\}$ be the parameters representing an
infinitesimal $SU(n)\times SU(M)$ gauge transformations and let
$\{\widetilde\theta_b\}$ be the rest of the gauge parameters such
that $\{\theta_a,\widetilde\theta_b\}$ represents a general infinitesimal
$SU(N)$ gauge transformation. The Faddeev-Popov ghost term in
\gfixterms\ represents the determinant coming from the measure in
changing coordinates on the gauge orbits from
$\{\theta_a,\widetilde\theta_b\}$ to $\{\theta_a,G_b\}$. Therefore,
even though ${\delta G^b\over\delta\theta^a}\ne 0$, only ${\delta
G^b\over\delta\widetilde\theta^c}$ contributes to the determinant.
The resulting Faddeev-Popov ghosts are in the bifundamental of
$SU(n)\times SU(M)$ only. To summarize, by choosing a gauge-fixing
function $G$ such that $G=[v,[v,\tilde G]]$, we have fixed the gauge
symmetry only
partly; the $SU(n)\times SU(M)$ subgroup is not fixed.

The Lagrangian decomposes as
\eqn\lagrang{\LL=\LL_{ga}+\LL_{gf}+\LL_{gh}~.}
Explicitly these are given by
\eqn\pieces{\eqalign{\LL_{ga}=&{1\over 4g^2}\tr F^2+{1\over
2g^2}\tr\[\bw\cdot\(D^2\)\bw+\(\bD\cdot\bw\)^2-i2w^\mu\[F_{\mu\nu},w^\nu\]+[w_\mu
,w_\nu][w^\mu ,w^\nu ]\]\cr\LL_{gf}=&-{1\over
2g^2}\tr\(\bD\cdot\bw\)^2\cr \LL_{gh}=&-{1\over g^2}\tr\[\bar c
D^2c\]~,}}
where $F$ and $\bD$ are with respect to the $SU(n)\times SU(M)$
fields ($\bA+\ba$). We now add two antisymmetric auxiliary fields
$d^{\mu\nu}$ and $e^{\mu\nu}$ transforming in the adjoint of $SU(n)$
and $SU(M)$ correspondingly. We use these to express the four $w$
interaction as in \fourvertex. In the resulting action, all the
dependence on the fields in the bi-fundamental of $SU(n)\times
SU(M)$ ($\bw$ and $c$) is quadratic. Integrating them out leads to the
following determinants
\eqn\dets{\eqalign{\bw:\quad &\det\[-D^2\eta+i2F+d+e\]^{-\half} \cr
c:\quad &\det\[-D^2\]^1~,}}
where for example $F=F_{\mu\nu}^af^{abc}$.

Next, we would like to add an adjoint Higgs field
$\widetilde\Phi$. It decomposes into a piece that does not commute
with $v$ and a piece that does, according to:
$$\chi=[v,[v,\widetilde\Phi]]~,\quad\varphi=\widetilde\Phi-\chi~.$$
The resulting theory is obtained by dimensional reduction from
the pure YM case
described above in 4+1 dimensions, where
$$\varphi=A_4+a_4\quad{\rm and}\quad\chi=w_4~.$$
The gauge fixing function that is obtained from \gaugefix\ in 4+1
dimensions is:\foot{This is called the background 't
Hooft-Feynman gauge in \StrasslerZR.}
$$G=\bD\cdot\bw-i\[\varphi,\chi\]~,$$
We then consider the theory in a state where
$$\<\varphi\>=mv~.$$
In blocks, $\varphi$ decomposes as
$$\varphi=mv+\(\matrix{\phi_{n\times n}& 0 \cr 0 &\Phi_{M\times M}}\)~.$$
The reduction of the determinant \dets\ in 4+1 to 3+1 dimensions is:
\eqn\wchidet{\eqalign{w,~\chi:\quad&\det\[\(\matrix{\(-D^2+\varphi^2\)\eta_{\mu\nu}
+2iF_{\mu\nu}+d_{\mu\nu}+e_{\mu\nu} &
2iD_\mu\varphi+d_{4\nu}+e_{4\nu}\cr\cr
-2iD_\mu\varphi+d_{\mu4}+e_{\mu4} & -D^2+\varphi^2}\)\]^{-\half}\cr
c:\quad&\det\[-D^2+\varphi^2\]^1~,}}
where for example $\varphi^2=\varphi^a\varphi^bf^{abc}f^{cde}$.
Next, we will express \wchidet\ in the worldline formalism.

\subsec{Worldline representation of determinants}

Next, to evaluate the planar quantum effective action \qea, we would
like to express
\eqn\logdet{\log\det\({\delta^2 S\over\delta^2
w}\)_{w=0}[\bA+\ba]=\log\det\[-D^2\]-\half\log\det\[-D^2\eta+2iF-d-e\]}
in the worldline formalism. The Higgsed theory is then obtained by
dimensional reduction as above. For $d=e=0$ the corresponding
worldline expression of \logdet\ was derived in \StrasslerZR\ (see
\Schubert\ for a detailed review). Since $d$ and $e$ are
anti-symmetric tensors, they can be regarded as corrections to $F$
and the same worldline expression with $2iF\to (2iF-d-e)$ applies.
If instead of introducing two anti-symmetric auxiliary tensor fields
($e$ and $d$) we had introduced a single auxiliary tensor
field in $SU(n)$ that is not anti-symmetric,
then in the resulting
1PI vertex we would have only auxiliary vertex insertions but no
auxiliary background fields. Here, for simplicity of presentation we have chosen to
introduce two anti-symmetric fields.

Below, we summarize the result of \StrasslerZR\ for \logdet. We
refer the reader to \refs{\StrasslerZR,\Schubert} for further
details. In Appendix \DDD\ we give a superspace expression for the
vector determinant of \StrasslerZR\ (at $d=e=0$).

The first term in \logdet\ is minus the contribution of a scalar
field. A worldline representation of it is\foot{Note that also the
gauge indices can be represented on the Hilbert space of worldline
fields \BalachandranYA\ whose string theory boundary counterparts
carry the Chan Paton indices \refs{\MarcusCM,\HellermanBU} (see
Appendix \DDD).}
$$\log\det\[-D^2\] =-\int_0^\infty{dT\over T}\NN\int [D\bx(\cdot)]\tr
Pe^{i\int_0^T\[\half\dot\bx^2+\bA\cdot\dot\bx\]}~,$$

The second term in \logdet\ is the contribution of the w-boson.
Following \refs{\StrasslerZR, \Schubert},
a
worldline representation of it is
\eqn\finaly{\eqalign{-&\half\Tr\log\[-D^2 +i2F-d-e\]\cr
=&\lim_{M\to\infty}\int_0^\infty{dT\over T}\NN\int [D\bx(\cdot)][
D\psi D\bar\psi]_{{\rm GSO}}\tr Pe^{i\int_0^T\half\[\dot\bx^2+
i\bar\psi^\mu\dot\psi_\mu+ 2\bA\cdot\dot\bx +
E_{\mu\nu}\psi^\mu\bar\psi^\nu\]-M^2T}~,}}
where
$$E_{\mu\nu}=-2iF_{\mu\nu} +
d_{\mu\nu} + e_{\mu\nu} + i M^2 \eta_{\mu\nu}$$
and $\psi^\mu$ is a complex fermion. The trace on the LHS \finaly\
is a color trace and a trace over lorentz indices. The subscript
`GSO' indicates that a sum over periodic and antiperiodic boundary
conditions on the worldline fermions should be performed; this
removes the states with even fermion number. The mass term $ M^2
\(\eta_{\mu\nu}\bar\psi^\mu\psi^\nu-1\)$ is added to remove from the
accessible spectrum the odd forms with fermion number larger than
one.

We will ignore the auxiliary fields for the rest of this subsection\foot{Note that the
auxiliary field $d\in a$ is set to zero at the end, whereas
integrating out $e\in A$ leads to a non-trivial contribution only to
loops with self-crossing points.}.

It
may be possible to write the combination of vector (including
unphysical modes) and ghost determinants as
\eqn\noghost{\Gamma[A]=\int_0^\infty{dT\over T}\NN\int
[D\bx][D\psi][D\bar\psi]\tr_{{\rm phys}}
Pe^{i\int_0^T\half\[\dot\bx^2+
i\bar\psi^\mu\dot\psi_\mu+2\bA\cdot\dot\bx-2iF_{\mu\nu}\psi^\mu\bar\psi^\nu\]}~,
}
where the trace is taken over the Hilbert space of physical states.
The worldline time evolution maps this space to itself. We have not
verified this equality for $A\neq 0$.

We will also be interested in the worldline theory for the theory
with an adjoint Higgs scalar. To add a worldline mass and a Higgs
field to the physical spectrum, we simply go to five dimensions and
fix the momentum in the fifth dimension to $p_4=m$. The fifth
worldline fermion $\psi^4$ remains and is necessary for worldline
supersymmetry. The fifth component of the background gauge field
becomes the Higgs field $A_4=\Phi$. The resulting worldline
representation of the YM + Higgs one loop effective action is
\eqn\HiggsYM{\eqalign{\Gamma[\bA,\Phi]=&-\half\log\det\[i\(\matrix{\(-D^2+m^2+g^2\Phi^2\)\eta_{\mu\nu}+2iF_{\mu\nu}
& 2igD_\mu\Phi\cr\cr -2igD_\mu\Phi & -D^2+m^2+\Phi^2}\)\]\cr
+&\qquad\log\det\[i(-D^2+m^2+\Phi^2)\]\cr\cr =&\int_0^\infty{dT\over
T}\NN\int [D\bx(\cdot)][D\psi D\bar\psi]\tr_{{\rm phys}}
P
e^{iS[\bx,\psi,\bar\psi;A,\Phi]}
~,}}
where
$$\eqalign{S=\int_0^T\{\half[\dot\bx^2-&m^2-\Phi^2]+
i\bar\psi^\mu\dot\psi_\mu-i\psi^\mu\bar\psi^\nu
F_{\mu\nu}+\bA\cdot\dot\bx\cr + &i\bar\psi^4\dot\psi_4
-i(\psi^\mu\bar\psi^4-\psi^4\bar\psi^\mu)D_\mu\Phi\}.}$$

\subsubsec{Background gauginos (or, turning on RR background fields)}

It is also interesting to consider gauge theories with fermionic
fields in the adjoint, such as the gauginos of supersymmetric gauge
theories. The components of these fields with the gauge quantum
numbers of $w$ are simple to include, by adding in their worldline
contribution: \eqn\spinoral{\eqalign{\Gamma_{\rm gaugino}[\bA]
=&\log\[\det\(\D+m\)\]=\half\log\[\det\(\D+m\)\(-\D+m\)\]\cr
=&\half\tr\log\((-D^2+m^2)\One-{i\over
4}F_{\mu\nu}[\gamma^\mu,\gamma^\nu]\)~.}}
%
Similarly to the scalar case, we
rewrite the fermionic determinant as
\eqn\ferworldl{\eqalign{ &\int_0^\infty{dT\over T}\NN\int
[D\bx][D\psi]\tr Pe^{i\int_0^T\half\[\dot\bx^2+2i\bA\cdot\dot\bx-
m^2+i\psi^\mu\dot\psi_\mu+i F_{\mu\nu}\psi^\mu\psi^\nu\]}~,}}
where the $\gamma$ matrices are represented by real worldline
fermions ($\psi$).

To turn on background value for the gauginos ($\Psi$), however, one
would like to add ``spin fields'' into the worldline formalism. We
then expect the Ramond and NS action given above to be realized in
different subspaces of the Hilbert space created by the spin fields.
We also expect the variation of the worldline action with respect to
the gaugino to gives the worldline gaugino vertex operator. It is
not clear if the worldline ghosts will still decouple. A
Green-Schwarz or Berkovits-like formulation of the worldline theory
(\eg\ \refs{\SorokinNJ}) would be better for this purpose (however,
it is not known how to couple these to an off-shell background).

\newsec{Discussion}

Like many of the results arising from worldline formalisms,
our prescription
has both perturbative and nonperturbative aspects. For example, we
must sum over all tree-level diagrams in the $a$s to retain gauge
invariance, just like at any order in the perturbation expansion in
a gauge theory. On the other hand, the decoupling of longitudinal
polarizations follows from integration by parts, like in string
theory. At strong coupling, we have tried to argue that the formula
becomes more string-like, since the one-particle-irreducible
contributions dominate.


In \AldayHR,
AdS was shown to be self-T-dual;
this suggested that the system should
also have the conformal symmetry of
the {\it dual} AdS space.
Since the formulae derived in this paper
give an interpretation of this T-duality
as a Fourier transform in loop space,
we had hoped that the formula would shed
light on the mysterious `dual conformal invariance'.

There are many approximate descriptions
of scattering processes that use Wilson loops.
Prominent among these are
heavy quark effective field theory (\eg\ \refs{\heavyquark}),
and the eikonal approximation
(\eg\ \refs{\eikonalrefs, \KovnerVI}).
The basic idea of both is that
a charged particle with enough inertia
moves in a straight line, and
its interactions 
are encoded entirely
in the phase it acquires in moving through the gauge field.
When applicable, a saddle point approximation
to our formula should reproduce these approximations.
The formula may be able to suggest
a nonperturbative formulation
of the inclusion of recoil corrections
to these approximations;
such a formulation would be useful,
for example in the study of jet quenching
at strong coupling \LRW.

\bigskip
\noindent
\centerline{{\bf Acknowledgements}}
\bigskip

We thank H.\ Elvang, A.\ Lawrence, H.\ Liu, J.\ Maldacena, H. Schnitzer, L.\
Susskind and C.-I\ Tan for useful conversations. We are especially grateful
to A.\ Polyakov for inspiring discussions. The work of J.M. is
supported in part by funds provided by the U.S. Department of Energy
(D.O.E.) under cooperative research agreement DE-FG0205ER41360. The
work of A.S. is supported in part by the DOE Grant No.
DE-FG02-92ER40706, and by an Outstanding Junior Investigator award.

\appendix{\CCC}{Reparametrization-invariant worldline theories}

The worldline theories described in \StrasslerZR\ are presented in a
way that looks like the string worldsheet in conformal gauge.
In addition, the vector determinant \logdet\
is expressed through as a sum of two worldline path integrals, one
for the w-bosons and the other for the ghosts. One would expect the
sum of these two worldline theories to result from gauge fixing
a single manifestly super-reparametrization invariant worldline
theory.\foot{Note that only the sum of these two worldline theories
is independent of the spacetime gauge we used in section 6 to obtain
the one loop determinants.} For the scalar and the spinor worldline
theories, a reparametrization and super-reparametrization invariant
description is indeed known \refs{\BrinkUF, \AndreevBZ}.
These theories are obtained from the bosonic string and the
Ramond sector of the superstring, respectively, on the annulus in the $\alpha'\to
0$ limit. In that limit all the stringy excitation decouple and the
string worldsheet theory on the annulus becomes a worldline theory on the
circle. Coupling the resulting worldline theory to an arbitrary
background gauge field without breaking the gauge symmetries is
straightforward. In these descriptions, after fixing a gauge,
unphysical modes are removed by constraints. For the worldline
description of the vector determinant \logdet, one would expect to
obtain such super-reparametrization invariant description from the
$\alpha'\to 0$ limit of the superstring in the Neveu-Schwartz sector.
However, such a description is not known. Our attempts at
formulating such a manifestly super-reparametrization invariant
description of the NS sector were not successful; we include them as
a cautionary tale for the reader. We did however find a superspace
description of the `NS sector' worldline, which may be useful and is
described in appendix \DDD.

\subsec{Gauge-fixed vector worldline}

\lref\HoweFT{
  P.~S.~Howe, S.~Penati, M.~Pernici and P.~K.~Townsend,
  Phys.\ Lett.\  B {\bf 215}, 555 (1988).
}

\lref\HoweVN{
  P.~S.~Howe, S.~Penati, M.~Pernici and P.~K.~Townsend,
  Class.\ Quant.\ Grav.\  {\bf 6}, 1125 (1989).
}

A full locally-symmetric description is actually not necessary
to obtain
a single worldline path integral representation of the vector
determinant. To see this, note that if we start with a worldline
theory consist of a vector of bosons ($x^\mu$) and a vector
of complex
fermions ($\psi^\mu$) with an action
$$S=\half\int d\tau\[\dot x_\mu\dot
x^\mu+i\bar\psi^\mu\dot\psi_\mu\]~,$$
then the operators
$$Q=p_\mu\psi^\mu~,\quad \bar Q=p_\mu\bar\psi^\mu~,
\quad \HH=\half\bp^2$$
are
conserved charges which
generate a global $\NN=2$ symmetry.
If at $\tau=0$ we start with a physical state, then the
state as well as the notion of physical does not change under
worldline time evolution. That is, if at time $\tau=0$ the state
$|\varphi\>$ is annihilated by $\HH$ and $Q$, it will continue to be
upon time evolution. In addition, since $[\HH,Q]=0$, the physical
constraint at time $\tau=0$ is not too restrictive, so the resulting
physical spectrum is not empty.

At this point this observation seems trivial, since there does exist
a locally super-reparametrization invariant description of the
theory (which is described in the next subsection). However, in the presence of
a background gauge field we do not know how to generalize that
theory such that it retains its local symmetries \HoweVN, and
our only consistent description of it will be as above.

The spectrum of this theory, after GSO projection,\foot{The GSO
projection is
implemented by a sum over periodic and antiperiodic boundary
conditions for the worldline fermions.} contains potentials of all
odd degree. To project out the unwanted three form, Strassler
\StrasslerZR\ added to the action the mass term
$$S_M=M^2 \int d\tau \(\psi_\mu\bar\psi^\mu-1\)~.$$
It affects only the three-form state, which is projected out in the
limit
$M^2\to\infty$. In the presence of $S_M$, the charge $Q$ is no
longer conserved. However, since
$$[\HH,Q]=M^2Q~,$$
when restricted to the kernel of $\HH$ and $Q$, under time
evolution, physical states remain physical and the physical spectrum
is not empty.

Next, to couple the theory to a background gauge field ($A$), we add
to the path integral the term
\eqn\Aterm{P e^{iS_A}~,}
where $P$ stands for path ordering and $S_A$ is
\eqn\SA{S_A=\int d\tau\[\dot x_\mu A^\mu-{i\over
2}\psi^\mu\bar\psi^\nu F_{\mu\nu}\]~.}
The second term in \SA\ is the supersymmetrization of the first. The
resulting Hamiltonian and supercharges are
$$\HH=\half\pi_\mu\pi^\mu~,\quad Q=\psi^\mu\pi_\mu~,\quad\bar
Q=\bar\psi^\mu\pi_\mu~,$$
where
$$\pi_\mu=p_\mu-A_\mu=\dot x^\mu~.$$
Since these close on the same algebra, when restricted to physical
states at $\tau=0$, the theory remains consistent. For non-abelian
gauge theories the $[A_\mu,A_\nu]$ part of $F_{\mu\nu}$ in $S_A$
seems to break worldline supersymmetry. However, the path ordered
exponent of the worldline action is supersymetric \AndreevBZ. To see
this note that the SUSY variation of the terms in the worldline
action linearly coupled to the external gauge field leads to a total
derivative ($\d_\tau [(\bar\epsilon\psi^\mu+\epsilon\bar\psi^\mu)
A_\mu]$). When expanding the exponent of the worldline action, these
gives boundary terms due to the path ordering
($(\bar\epsilon\psi^\mu+\epsilon\bar\psi^\mu) [A_\mu,A_\nu] \dot
x^\nu$). These terms exactly cancel the SUSY variation of
$\psi^\mu\bar\psi^\nu\[A_\mu,A_\nu\]$ coming from one lower
order in the expansion.

\subsec{Local super-reparametrization symmetry without background
gauge field}

In this subsection we describe a local $\NN=2$
super-reparametrization invariant worldline theory without
background gauge field.

Consider the first order super-reparametrization invariant action
\eqn\act{S=\int d\tau\[p_\mu\dot x^\mu+{i\over
2}\psi^\mu_i\dot\psi_\mu^i-e\HH-i\chi_i\QQ_i\]~,}
where
$$\HH=\half p_\mu
p^\mu~,\quad\QQ_i=p_\mu\psi_i^\mu~,\quad i=1,2~.$$
Here $e$, $\chi_1$ and $\chi_2$ are gauge fields gauging the local
symmetry generated by
$$ G=\alpha\HH+i\epsilon_i\QQ_i$$
under which the fields transform as
\eqn\susy{\eqalign{&\delta x^\mu=\alpha
p^\mu+i\epsilon_i\psi^\mu_i\cr &\delta p^\mu=0\cr
&\delta\psi_i^\mu=-\epsilon_i p^\mu\cr &\delta
e=\dot\alpha-2i\epsilon_i\chi_i\cr &\delta\chi_i=\dot\epsilon_i~.}}
Integrating out $p^\mu$ amounts to plugging in its equation of
motion
%
\eqn\pval{p^\mu={1\over e}\(\dot x^\mu-i\chi_i\psi_i^\mu\)~.}
The resulting action is
\eqn\secondorder{S=\int d\tau{1\over 2e}\[\(\dot
x_\mu-i\chi_i\psi_i^\mu\)^2+ie\psi_i^\mu\dot\psi_\mu^i\]~.}
Upon quantization, the fields satisfy the canonical commutation
relations
$$[x^\mu,p^\nu]=i\eta^{\mu\nu}~,\quad\{\psi_i^\mu,\psi_j^\nu\}=\eta^{\mu\nu}
\delta_{ij}~.$$
We defined
$$\eqalign{\psi^\mu=&\psi_1^\mu+i\psi_2^\mu~,\cr
\bar\psi^\mu=&\psi_1^\mu-i\psi_2^\mu~,}\qquad\eqalign{Q=&\QQ_1+i\QQ_2=p_\mu\
\psi^\mu\cr\bar
Q=&\QQ_1-i\QQ_2=p_\mu\bar\psi^\mu}~.$$
These satisfy the commutation relations
$$\{Q,\bar Q\}=2\HH~,\quad \[\HH,Q\]=\[\HH,\bar Q\]=0~.$$
Next, we quantize the theory in the Gupta-Bleuler style by imposing
that physical states are annihilated by $\bar Q$ and $\HH$. At each
momentum $p$, the Hilbert space is spanned by the states
$$\eqalign{&|0,p\>~,\quad\varepsilon_\mu\psi^\mu|0,p\>~,\quad
f_{\mu\nu}\psi^\mu\psi^\nu|0,p\>\cr
&g_{\mu\nu\rho}\psi^\mu\psi^\nu\psi^\rho|0,p\>~,\quad
\psi^0\psi^1\psi^2\psi^3|0,p\>~.}$$
By GSO projection (\ie\ gauging the worldline fermion number), we
project to the states
$$\varepsilon_\mu\psi^\mu|0,p\>\quad{\rm and}\quad
g_{\mu\nu\rho}\psi^\mu\psi^\nu\psi^\rho|0,p\>$$
only. The Hamiltonian constraint is
$$p^2=0$$
and the $\bar Q$ constraint reads
$$p^\mu\varepsilon_\mu=0~,\quad p^\mu g_{\mu\nu\rho}=0~.$$
The remaining states are a transverse one form and transverse three
form. After the
physical projection, there are no negative norm states in the
physical spectrum.

Now consider the addition of $S_A$, the coupling to
the background gauge field.
With the supersymmetry transformations \susy,
this term is not invariant.
To make this term invariant
will require some nonlinear realization of the
worldline local supersymmetry, which
is beyond the scope of this paper.

\appendix{\DDD}{String-inspired and superspace
worldline theories}

The worldline theories in question can in principle be obtained in a
very direct way from string theory: they describe the zero-slope limit of an
open type IIB superstring stretched between two parallel D3-branes.
Using the RNS formalism for this string worldsheet in conformal
gauge, we obtain the spinors from the R sector and the vectors and
scalars from the NS sector. Finally, we give a superspace
representation of the vector (NS) worldline theory coupled to
background gauge field.
We ignore the auxiliary fields introduced in sections 4 and 6 for
this discussion.

\subsec{Dimensional reduction of the open string}

In the NS sector of an open superstring,
on the doubled strip ($\sigma \in [0, 2\pi)$),
the fermions have boundary conditions
$$ \psi^\mu( \sigma^1 + 2\pi , \sigma^2 ) = -
\psi^\mu( \sigma^1, \sigma^2 )~  .$$
The general field with such boundary condition can be expanded as
$$ \psi^\mu( \sigma^1, \sigma^2 ) =
\sum_{r \in \IZ+\half} \psi_r(\sigma^2) e^{ i r \sigma^1}~ ; $$
note that we have not solved the equations of motion, only the
boundary condition. Plugging into the action, this gives
$$ S[\psi] =
\int d\sigma^1 d\sigma^2 ~~
\half \eta_{\mu\nu} i \psi^\mu \( \del_2 - \del_1 \) \psi^\nu
$$
$$
= \int d\sigma^2 \sum_{r \in \IZ+\half} \half
\psi_r \( i \del_2 + r \) \psi_{-r}
 = \int d\sigma^2 \sum_{r =\half, {3\over 2} ... > 0}
\psi_r \( i \del_2 + r \) \psi_{-r}~.
$$
We will truncate this set of modes to
only the lowest-lying conjugate pair $r = \pm \half$;
only these modes create massless states.
The masses of the other states are larger by
an amount proportional to the string scale
which we will take to $\infty$.

The worldsheet gravitino has a spin structure
which is correlated with that of the RNS fermions:
$$ \chi( \sigma^1 + 2\pi , \sigma^2 ) = -
\chi( \sigma^1, \sigma^2 )~.$$
Reducing the coupling to the
supercurrent gives
$$ \int d\sigma^1 d\sigma^2~ \chi T_F
= \int d\sigma^2 \( \chi_{\half} G_{-\half} + \chi_{-\half}
G_{\half} + \dots \)~. $$

The worldline action
(in units where $ { 2\pi \over 4 \pi \alpha'} = \half $ )
is
$$
S = \int dt \( \half \dot \bx^2 + \bar \psi_\mu \( i \del_t + \mu \)
\psi^\mu \)~,
$$
where we have changed the name of the worldline time $ \sigma^2 = t
$ and $ \psi \equiv \psi_{-\half}, \bar \psi \equiv \psi_{\half}$,
and $ \mu = \half$. Notice that supersymmetry is `broken,' in the
sense that bosonic and fermionic states related by the action of $G$
do not have the same energies -- of course, this is a consequence of
the NS boundary condition. The constraint algebra, however, still
closes and is generated by
$$ G_{-\half} \equiv Q = \psi^\mu p_\mu, ~~~~
G_{\half} \equiv \bar Q = \bar \psi^\mu p_\mu $$
$$ H = \half p^2 + \mu J, $$
where $ J  \equiv  \psi \bar \psi $ is the fermion number operator
$ [J, \psi] = \psi, [J, \bar \psi] = - \bar \psi$.
The worldline algebra (before coupling to a background gauge field)
is
\eqn\QQH{
 \half \{ Q, \bar Q\} = H - \mu J, ~~Q^2 = 0 = \bar Q^2 , ~~ [ H, Q]
=\mu Q , ~~ [ H, \bar Q ] = - \mu \bar Q ~.}

In the same way that in old covariant quantization of the
superstring,
the modes of the supervirasoro algebra
are treated as Gupta-Bleuler constraints,
$$ G_{r > 0} \ket{{\rm phys}} = 0 $$
(since this is enough to guarantee that all physical-state matrix
elements
of the constraint algebra generators vanish),
we should only impose that
$$ \bar Q \equiv G_{\half} $$
annihilate physical states.

In the superstring, the vacuum energy must take a definite value
to be consistent with conformal invariance.
This value is such that states with a single $\psi_{-\half}$
excitation
are massless.  We impose this value of the vacuum energy in our
worldline theory.

The worldline theory also inherits the GSO projection onto states
with one parity of the fermion number.
That is the parity such that the massless states survive.
Note that the worldsheet ghosts carry GSO-charge;
if we were careful about them in the worldline theory
(which would clearly be best done simply by dimensionally reducing
the NS sector ghosts of the superstring $b_0, c_0, \gamma_{\pm
\half}, \beta_{\pm \half}$),
we could see that the NS ground state has odd fermion number.

If we perform the same dimensional reduction in the Ramond sector,
we obtain the usual spinning particle of \BrinkUF. The GSO
projection, consistent in even numbers of dimensions, makes the
resulting spinor chiral.

\subsubsec{Adding a spacetime mass}

So far we have discussed the dynamics of the coordinates {\it along}
a Dp brane (and therefore $\mu=0,\dots,p$). The directions
perpendicular to the brane are obtained from the above simply by
setting their momentum $p_i=0$, $A_i=0$ where $i=p+1,\dots,d$. Of
course, the dimensions we are most interested in are $p=3$ and
$d=9$. However, the worldline theory seems to be consistent for any
number of dimensions.

Next, we would like to Higgs the theory. From the point of view of
the string theory, that can be accomplished by separating the
D-branes. We can then redo the dimensional reduction in a sector
with nonzero winding: \eqn\windingx{ x^4(\sigma, t) = \Delta y
\sigma } but again eliminate all oscillator modes. In the R-sector,
this procedure reproduces the massive spinning particle \BrinkUF.
The mass $m$ appearing there is
\eqn\massfromdeltay{m={\Delta y \over \alpha'}}
and is kept fixed in the zero slope $\alpha'\to 0$ limit. In the NS
sector, the resulting supercharges (ignoring the terms involving CP
fields which are unaffected) are
\eqn\massiveQs{Q=\psi_\mu \dot x^\mu + m \psi^4~,\quad \bar Q=\bar
\psi_\mu\dot x^\mu+m\bar\psi^4~,}
where $\psi^4 \equiv \psi^4_{-\half}, \bar \psi^4 \equiv \psi^4_{
\half}$.

Finally, we must discuss the Faddeev-Popov ghosts for the vector case. They
again decouple from the worldline action of the matter fields
($x,\psi, \eta$).  Their partition function is of the form \eqn\vectorghosts{
Z_{gh} = \int [Db_0 Dc_0
D\beta_{\half}D\beta_{-\half}D\gamma_{\half}D\gamma_{-\half}] ~~
e^{i S_{{\rm gh}}} .}

\subsec{Superspace representation of the worldline theory}

Having a superspace representation of a supersymmetric theory is a
very useful technical tool. The spinor worldline theory has an
$\NN=1$ superspace description \BrinkUF. Similarly, one would like
to have an $\NN=2$ superspace description of the vector theory
described in \StrasslerZR\ and summarized in section 6. However, the
presence of a mass for the worldline fermions manifestly break the
global $\NN=2$ supersymmetry. Here we give a deformed $\NN=2$
superspace representation of it. The ability to write down such a
superspace description is related to the fact that at the level of
the action (but not the path integral measure) a mass ($\mu$) for
the worldline fermions can be absorbed in a field redefinition
$\psi\to e^{i\mu t}\psi$.

Define
      $$D_0 \equiv \del_\theta + \bar \theta i \del_t, ~~~~~~
      \bar D_0 \equiv \del_{\bar \theta} + \theta  i \del_t~ ;$$
these are the $\CN=2$ superspace derivatives with $\mu=0$, \ie\ the
      ordinary ones.
$\CN=2$ superspace derivatives
which generate the truncated NS algebra are
         $$ D = e^{i \mu t} \(D_0 + \mu \theta \bar \theta
\del_\theta\)
      ~~~~~~ \bar D = e^{-i\mu t}\( \bar D_0 + \mu \theta \bar \theta
\del_{\bar \theta}\) .
       $$
       They satisfy
      $$ D^2 =0
      , ~~~~~~
      \bar D^2 =0
       $$
$$ \{D, \bar D\} = 2 i \del_t  - 2 \mu\(\theta \d_\theta - \bar
\theta \d_{\bar\theta}\)
= -2 \(H+  \mu  J\)~ .
$$
The fact that $D^2=0 $ is made manifest by the observation that
$D$ can be rewritten as
$$ D = e^{ i \mu \bar y} D_0 , ~~~~ \bar D = e^{ -i \mu y} \bar D_0~,
$$
where $y$ is the chiral time coordinate
$$ y \equiv t + i \theta \bar \theta $$
which satisfies $ \bar D y = 0 $ (and $ \bar D_0 y = 0$).

Each of the worldline supercoordinates $X^\mu$ comprises a real
superfield:
      $$ X  = x + \theta \psi + \bar \theta \bar \psi + \theta
      \bar \theta
      F~ . $$

Also, we will need to be more explicit
about the gauge representation of the worldline.
To do this it is useful to introduce worldline degrees of freedom
$\eta$
whose hilbert space generates the Chan-Paton space
\BalachandranYA;
from the point of view of the zero-slope open string, these are
boundary degrees of freedom \refs{\MarcusCM,\HellermanBU}.
We will follow Friedan and Windey \FriedanXR.

\def\underscore{\underline}
The modes generating the CP $\space$
   $\underscore{\eta}^{a=1..N}$ will be chiral superfields
   whose lowest components are fermions (fermi multiplet):
      $$ \bar D \underscore{\eta} = 0  ~~~\Longrightarrow
~~~ \underscore{\eta} = \eta(y)+ \theta b(y) $$
      where the components are functions of
      the chiral time coordinate $y$.

      A reasonable lagrangian which contains kinetic terms
for $X$ and $\eta$ is
\eqn\starequation{
       \half\int d^2\theta ( D X \bar D X + \bar \eta \eta )
      =
      \half \dot x^2 + i \bar \psi \dot \psi -\mu \bar \psi \psi
+{F^2\over 2}
            + i \bar \eta \dot \eta- \half\bar b b~;}
here we define $ \int d^2 \theta \equiv \half [\bar D,D] $.
      The similarity between the fields generating the CP space
      and the worldline fermions has often been remarked upon;
it is a worldline remnant of the fact that their 2d avatars both
generate an $SU(N)$ current algebra.
The supercharges are
\eqn\superchargeswithoutA{Q=\psi\dot x+i\bar\eta b~,\qquad\bar Q=
\bar\psi\dot x-i\eta\bar b~.}

The states of the $\eta$'s generate $2^N=\oplus_{p=0}^N\wedge^p{\bf
N}$ states. We will be interested mainly in the case where the
worldline transforms in the fundamental ${\bf N}$ of the $U(N)$
group. To accomplish the projection onto this irrep, we introduce a
worldline gauge field $a_0$, which is a supersymmetry singlet. We
add a 1d Chern-Simons term $\Delta S=q\int dt\,a_0$ and couple $a_0$
to the $\eta$ number current.  The terms involving $a_0$ are then
\eqn\gaugeetanumber{\int dt\,a_0\(\eta\bar\eta-1\)~.}
This action is supersymmetric on-shell (\ie\ after using the
equations of motion for $b$)\foot{ An off-shell supersymmetric
version of the first term in \gaugeetanumber\ is
$$ \int d^2 \theta\int dt\,\bar\eta(t)\,a_0(t)\int^t
dt'\,\eta(t')~.$$
In checking this, it is convenient to choose one-dimensional Lorentz
gauge: $ \del_t a_0 = 0 $.  Note, however, that this expression is
not gauge invariant. The one-dimensional Chern-Simons term $ q \int
dt a_0 $ is supersymmetric, and gauge invariant as long as $q \in
\IZ.$ }; this is because the $\eta$-number current commutes with the
supercharges \superchargeswithoutA\ on-shell. If for some reason one
want to generate a symmetric-tensor representation of $U(N)$, one
should introduce bosonic chiral multiplets on the worldline.

\subsubsec{Coupling to a background gauge field}

The superspace expression on the LHS of \starequation\ above
includes the coupling to the gauge field if one promotes the action
of $D$ on $\eta$ to the covariant one $D_A$ as in Windey et al. We
modify the superspace derivatives when they act on charged fields to
$$D \longrightarrow D_A = D + iDX^\mu A_\mu(X)~,\qquad
\bar D \longrightarrow \bar D_A = \bar D + i\bar DX^\mu A_\mu(X)~.$$
The algebra is
$$ \{ D_A, \bar D_A\} = \{ D + iDX A , \bar D + i\bar DX A \} $$
$$= i \del_t + \{ iDX A , \bar D\} + \{ D, i\bar DX A \} - \{ DX A ,
\bar DX A\} $$
$$ = i \del_t + i\( \{ \bar D, D\} X \) A + iDX \bar D A + i\bar D X
DA
- DX^\mu \bar DX^\nu[ A_\mu, A_\nu] $$
$$ = i \del_t - \del_t X^\mu A_\mu + iDX^\mu \bar D X^\nu
F_{\mu\nu}~.$$
Similarly
$$ D_A^2 = iDX^\mu DX^\nu F_{\mu\nu}~,\qquad
\bar D_A^2 = i\bar DX^\mu \bar DX^\nu F_{\mu\nu}~.$$

We now demand that $\eta$ is covariantly chiral: $ \bar D_A \eta
=0$. The kinetic terms for $\eta^{a=1..N}$ become
\eqn\etaint{\eqalign{\half\int d^2\theta \bar \eta_a \eta^a =&
-\half\bar D_A \bar \eta_a D_A \eta +{1\over 4} \( \{ D_A,\bar D_A\}
\bar \eta_a \) \eta^a -{1\over 4} \bar \eta_a \( \{ \bar D_A ,
D_A\}\eta^a\)\cr =& \bar \eta \( i \del_t - \dot x^\mu  A_\mu +
i\psi^\mu \bar \psi^\nu F_{\mu\nu} \) \eta-\half\bar b b~.}}
Since the NS mass ($\mu=\half$) affects only the higher components of
superfields, it has no effect on \etaint. Note that $\eta$ and $b$
appearing in the previous equation are defined as
$$\eta\equiv\underscore{\eta}|_{\theta =\bar \theta = 0 }~,\qquad
b \equiv e^{-i\mu t} D_A \underscore{\eta}|_{\theta =\bar \theta = 0
}. $$ Integrating out the auxiliary fields $b$ by their algebraic
equations of motion gives $ b = 0 = \bar b$. The covariant
supercharges expressed as differential operator in superspace is
\eqn\supercharge{\eqalign{Q_A=&e^{-i\mu
y}\(\d_\theta-i\bar\theta\d_t\)+i\[Q_0,X^\mu\]A_\mu(X)\cr \bar
Q_A=&e^{i\mu\bar y}\(\d_{\bar\theta}-i\theta\d_t\)+i\[\bar
Q_0,X^\mu\]A_\mu(X)~,}}
where $A$ is a color matrix acting only on $\underscore\eta$ and
$\mu$ act only on $X$. The Noether currents associated to the
corresponding supersymmetry transformations are\foot{For
convenience, we have rescaled the supercharges by a factor of $i$.}
\eqn\superchargemu{\eqalign{Q=&e^{-i\mu t}\[\psi^\mu \dot x_\mu
+i\bar\eta b\]=e^{-i\mu t}\[\psi^\mu p_\mu + i\bar\eta\(b+i\psi^\mu
A_\mu\eta\)\]\cr \bar Q= &e^{i\mu t}\[\bar \psi^\mu \dot
x_\mu+i\eta\bar b\]=e^{i\mu t}\[\bar\psi^\mu p_\mu +i\(\bar
b+i\bar\eta \bar\psi^\mu A_\mu\)\eta\]~.}}
Note that these are the non-covariant supercharges. Their
commutation relation with $H$ are also obtained from the Noether
theorem as
\eqn\Noether{\delta_{\epsilon Q} S=i\int dt\epsilon\[H, Q\]=i\int
dt\,\epsilon \mu Q~,} which is true for any epsilon and hence
reproduces \QQH.
Finally, we have
$$\{Q,\bar Q\} =\(\bp+\bar\eta\bA\eta\)^2+i\psi^\mu\bar\psi^\nu
F_{\mu\nu} =2\(H-\mu J\)~,$$
where $H$ is the operator that generates the non-gauge-covariant time
derivative ($\[H,\OO\]=-i\d_t\OO$) and we have set $b=0$. The
supercharges commute with the $\eta$-number current on-shell, so the
term \gaugeetanumber\ is supersymmetric.\foot{Since \gaugeetanumber\
is neutral, the covariant and the non-covariant supercharges act on
it in the same way.}

Note that it is the non-covariant hamiltonian and supercharges that
generate a symmetry
of the worldline action coupled to a fixed background gauge field.
Therefore,
it is these charges that are gauged on the worldline.
Once we include the path integral over the
background gauge field, also the covariant hamiltonian generates a
symmetry (which is gauged).

\appendix{\FFF}{Wilson sums and the loop operator}

In \PolyakovTJ, A. Polyakov suggested a relation between the gauge
theory loop operator for momentum Wilson loops and the dual string
theory Virasoro generator. The action of the loop operator on
Migdal's momentum Wilson loop for the polygon \momentuml\ is
\Migdal\foot{We remind the reader that $\widetilde
W[\bp(s)]_{Migdal}\ne\widetilde W[\bp(s)]$.}
\eqn\PolyakovmLoop{\hat L\widetilde
W[\bp(\cdot)]_{Migdal}=\sum_i\bk_i^2~\widetilde
W[\bp(\cdot)]_{Migdal}~.}
Polyakov interpreted the momentum Wilson loop evaluated on the
polygon as describing an open string stretched between D-instantons
(which are the T-dual to the D3-branes) located at the positions
$\{\bp_j=\sum_{i\le j}\bk_i\}$ and the loop operator as the
integrated Virasoro generator $L_0$ in the $\alpha'\to 0$
limit.\foot{The loop equation then equates \PolyakovmLoop\ to a sum
over divisions of $W[\bp(s)]_{Migdal}$ into two sub-loops.}

In the previous sections we have expressed scattering amplitudes
in terms of
Wilson sums. Next, to obtain an analog of \PolyakovmLoop, we will act
with the loop operator on the 1PI scalar vertex written in terms of a
Wilson sum.

The loop operator is a differential operator acting on functionals
of closed loops. One representation of it is
\eqn\loopoperator{\hat L\equiv\lim_{\epsilon\to 0}\oint
d\tau\int_{-\epsilon}^\epsilon ds
{\delta^2\over\delta\bx(\tau)\cdot\delta\bx(\tau+s)}~.}
Even though it is expressed in terms of some parametrization of the loop
$\bx(s)$, $\hat L$ is reparametrization invariant and therefore has
a well defined action on loop space functionals.

As defined in \loopoperator, $\hat L$ is the natural loop operator
that acts on Wilson loops with standard coupling to the background
gauge field
$$\oint\bA\cdot d\bx~.$$
These are the building blocks of the scalar contributions to the 1PI
vertex.
Without using the trick \spinfactor,
the $w$-boson contribution also has the coupling
$$\int ds\psi^\mu\bar\psi^\nu F_{\mu\nu}~.$$
The corresponding loop operator has additional terms involving the
worldline fermions \refs{\MigdalRN, \DrukkerZQ}. Here, for
simplicity, we consider the action of the loop operator on the
scalar 1PI vertex.

The scalar 1PI vertex $A_n^\bP$ \ampWilson, is not a functional of
position space loops but a function of the external momenta and
polarization. However, it is given in terms of a sum over closed
loops. Each closed loop in that sum is weighted by a product of
functionals. One is the Wilson loop expectation value
$\<W[\bx(\cdot)]\>$, another is the functional
$$F[\bx(\cdot)]\equiv\prod_{i=1}^n\int_{s_{i-1}}^T
ds_i~\varepsilon_i\cdot\dot\bx(s_i)~e^{i \bk_i\cdot\bx(s_i)}~~~,$$
and the rest can be attributed to the measure.

Next we act with the loop operator on the Wilson loop expectation
value and perform an integration by parts in the integral over
closed loops to obtain a relation analogous to
\PolyakovmLoop:
\eqn\lleinpol{\eqalign{&\int{dT\over T}\int \[D\bx(\cdot)\]
e^{-\int_0^T\half\dot\bx^2ds}
F[\bx(\cdot)] ~ \hat L\<W\[\bx(\cdot)\]\>\cr\cr =&\int{dT\over
T}\int \[D\bx(\cdot)\] e^{-\int_0^T\half\dot\bx^2ds}
F[\bx(\cdot)] ~
\<W\[\bx(\cdot)\]\>\sum_l\[\Delta\dot\bx(\tau_l)+i\Delta\bp(\tau_l)\]^2\cr
+&\int{dT\over T}\int \[D\bx(\cdot)\]
e^{-\int_0^T\half\dot\bx^2ds}\<W\[\bx(\cdot)\]\>\(\prod_{i=1}^n\int_{s_{i-1}}^T
ds_i~e^{i \bk_i\cdot\bx(s_i)}\)\sum_{j=1}^n\(\prod_{k\ne
j}\varepsilon_k\cdot\dot\bx(s_k)\)\cr
&\times\varepsilon_j\cdot[\Delta\dot\bx(s_j)+\Delta\bp(s_j)]\[\delta(s_j-s_{
j+1}) -\delta(s_j-s_{j-1}) -i\bk_j\cdot\dot\bx(s_j)\]~,}}
where for any wordline function $\by$,
$$\Delta\by(\tau)\equiv\by(\tau+0)-\by(\tau-0)$$
is the discontinuity of $\by(s)$ at $s=\tau$ and $\{\tau_l\}_l$ are
the points were $\dot\bx$ or $\bp$ have a discontinuity. $\bp(s)$ is
given in \momentuml.  With these definitions, $\Delta \bp$ is only
nonzero when $\tau = s_i$ is the location of a vertex insertion, in
which case $\Delta \bp = \bk_i$. In the case that vertices collide,
$s_i = s_{i+1}$, the jumps in momenta add; altogether
$$\Delta\bp(\tau)=\left\{\matrix{\bk_i&\tau=s_i\in(s_{i-1},s_{i+1})\cr
\bk_i+\bk_{i+1}& \tau=s_i=s_{i+1}\cr \vdots&\vdots\cr 0&{\rm
otherwise}}\right.$$

The differences between \lleinpol\ and \PolyakovmLoop\ come from the
difference in the measures (which gives the $\Delta\dot\bx$ terms)
and the polarization dependence of the gluon vertex
$\varepsilon\cdot\bx$ (which gives the last line in \lleinpol).  The
loop equation can be further used to express $\hat
L\<W[\bx(\cdot)]\>$ as a sum of the self-crossing points of
$\bx(\cdot)$ times the product of the two sub-loops (see for example
\Migdal).

\listrefs
\end